%% file: root.tex
\documentclass[12pt,letterpaper]{report}

\usepackage{geometry}
\usepackage{setspace}
\usepackage[font={footnotesize}]{caption}

\usepackage{amsmath}
\usepackage{amssymb}

\usepackage{bussproofs}
\usepackage{array}

\usepackage{cite}
\usepackage[hidelinks]{hyperref}

\newcommand{\abs}[2]{\lambda #1.#2}
\newcommand{\appTwo}[2]{#1\ #2}
\newcommand{\appThree}[3]{#1\ #2\ #3}
\newcommand{\appFour}[4]{#1\ #2\ #3\ #4}

\newcommand{\headPred}[1]{\mathcal{H}_p(#1)}

\newcommand{\ctx}[1]{\mathsf{ctx\_}#1}
\newcommand{\ctxmem}[1]{\ctx{#1}\mathsf{\_mem}}
\newcommand{\subctx}[2]{\ctx{#1}\mathsf{\_subctx\_}\ctx{#2}}
\newcommand{\init}{{\sl init}}
\newcommand{\focus}{{\sl focus}}

\begin{document}

\title{Automating the Proofs of Strengthening Lemmas in the Abella Proof Assistant}
\author{Dawn Michaelson}

\begin{titlepage} 
  \thispagestyle{empty}
  \fontsize{16pt}{16pt}

  \vspace*{1in}

  \begin{center}
    \textbf{Automating the Proofs of Strengthening \\Lemmas 
            in the Abella Proof Assistant}

    \vspace*{1in}

    \fontsize{14pt}{14pt}
    \textbf{Dawn Michaelson}
  \end{center}

  \vspace*{1in}

  \fontsize{12pt}{12pt}
  \noindent Submitted under the supervision of Gopalan Nadathur to the University Honors Program at the University of Minnesota-Twin Cities in partial fulfillment of the requirements for the degree of Bachelor of Science, \textit{summa cum laude} in Computer Science.

  \vspace*{1.5in}

  \begin{center}
    \textbf{April 27, 2017}
  \end{center}

\end{titlepage}

\pagenumbering{gobble}
\input{acknowledgements}

\input{abstract}

\tableofcontents

\pagenumbering{arabic}

\doublespacing


\input{introduction}

\input{hohh} 

\input{dependencies} 

\input{abella} 

\input{strengthening} 

\input{conclusion}

\bibliography{references}{}
\bibliographystyle{plain}

\end{document}

%% file: acknowledgements.tex
\chapter*{Acknowledgements}

{\doublespacing

I would like to first and foremost thank my advisor Gopalan Nadathur for his patience and teaching as I learned what I needed to know to do this work, and for his willingness to talk over any problems I was having with my work.  I appreciate all the time and effort he has put into helping me succeed in this.

I would also like to thank Yuting Wang, who helped me learn the background information I needed for this work while taking time out from writing his Ph.D. thesis to assist me.  This work was originally inspired by a paper he wrote with Kaustav Chaudhuri, and so I would like to thank both of them for that as well.

Thank you also to Professor Eric Van Wyk and Professor Nick Hopper for agreeing to be readers on my thesis.

This honors thesis is based upon work partially supported by the 
National Science Foundation under Grant No. CCF-1617771. Any opinions,
findings, and conclusions or recommendations expressed in this
material are those of the author and do not necessarily reflect the
views of the National Science Foundation.

}

%% file: abstract.tex
\chapter*{Abstract}

{\doublespacing
  
In logical reasoning, it is often the case that only some of a
collection of assumptions are needed to reach a conclusion. A
{\it strengthening lemma} is an assertion that a given conclusion
is independent in this sense of a particular
assumption. Strengthening lemmas underlie many useful techniques for 
simplifying proofs in automated and interactive theorem-provers. For
example, they underlie a mechanism called {\it subordination} that is useful
in determining that expressions of a particular type cannot contain
objects of another type and in thereby reducing the number of cases to be
considered in proving universally quantified statements. 

This thesis concerns the automation of the proofs of strengthening
lemmas in a specification logic called the logic of hereditary Harrop
formulas (HOHH). The Abella Proof Assistant embeds this logic in a
way that allows it to prove properties of both the logic itself and of
specifications written in it. Previous research has articulated a
(conservative) algorithm for checking if a claimed strengthening lemma
is, in fact, true. We provide here an implementation of this algorithm
within the setting of Abella. Moreover, we show how to generate an
actual proof of the strengthening lemma in Abella from the information
computed by the algorithm; such a proof serves as a more trustworthy
certificate of the correctness of the lemma than the algorithm
itself. The results of this work have been incorporated into the
Abella system in the form of a ``tactic command'' that can be invoked
within the interactive theorem-prover and that will result in an
elaboration of a proof of the lemma and its incorporation into the
collection of proven facts about a given specification. 



}

%% file: introduction.tex
\chapter{Introduction}

A strengthening lemma is a logical statement that any proof of a given
statement is independent of some of the available assumptions. Such
lemmas have many uses in the context of automated and interactive
theorem-provers. This thesis considers the generation of the proofs of
such lemmas without human assistance. The idea that underlies the
determination of the truth of such lemmas is to state: A conclusion
$C$ is independent of an assumption $F$ if an analysis of the
structure of proofs will show that $F$ cannot figure in a proof of
$C$. We describe how such an analysis can be carried out within a
specification logic called the logic of hereditary Harrop formulas
that is useful in formalizing rule-based descriptions of a variety of
computational systems. We also discuss how such an analysis can be
expanded into an actual proof of the relevant strengthening lemma.

It is perhaps useful to consider an example of the method we use
before we get into a detailed description of our specification logic 
and the automation of strengthening lemmas concerning it. Suppose that
we want to define a predicate called $\mathsf{list\_minus}$  that
holds between two lists and an element just in the case that the
second list is the result of removing the given element from the first
list. This relation can be defined by the following logical formulas:
\begin{center}
  \begin{tabular}{c}
    $\forall X \forall L\ \appFour{\mathsf{list\_minus}} {X} {(X::L)} {L}$ \\
    $\forall X \forall Y \forall L \forall L'\ (\appFour{\mathsf{list\_minus}} {X} {L} {L'}  \supset  \appFour{\mathsf{list\_minus}} {X} {(Y::L)} {(Y::L')})$
  \end{tabular}
\end{center}
It should be intuitively clear how a formula of the form 
$\appFour{\mathsf{list\_minus}} {X} {L_1} {L_2}$, where $X$, $L_1$, and
$L_2$ represent particular values, would have to be proved based on
this definition: we 
have either to show that our ``goal'' is an instance of the first
formula or that it matches with the righthand side of an instance of the
second formula and whose lefthand side is provable by a similar
process. This procedure is, in fact, the way in which derivations are
constructed in the logic of Horn clauses that underlies logic
programming languages like Prolog. 

Now, suppose that our assumption
set includes formulas for another predicate called $\mathsf{append}$ that
expresses the fact that three lists are in the ``append'' relation. 
Clearly, the procedure that we have described for proving
$\appFour{\mathsf{list\_minus}} {X} {L_1} {L_2}$ has no use for these
additional formulas and would therefore succeed or fail {\it
  independently} of their existence in the collection of
assumptions. Thus we can say that the assertion that a proof exists
for the given conclusion from an assumption set containing the definition of
$\mathsf{append}$ can be strengthened to an assertion that such a
proof exists even if the formulas for $\mathsf{append}$ are dropped. 

The idea that we have outlined above corresponds to a kind of
reachability analysis over formulas based on a derivation relation for
a given logic. An algorithm for conducting such a reachability
analysis for the HOHH logic has been described in \cite{yutingPaper}. This thesis
implements this algorithm and it then uses this implementation to
generate explicit proofs for the strengthening lemmas that are
validated by the algorithm. To prove such lemmas explicitly, it uses
the Abella Proof Assistant that encodes the HOHH logic and is capable
of proving meta-theorems about it. The end result of this work is a
new ``tactic command'' that can be invoked in the Abella system to
check and then automatically generate proofs for strengthening
lemmas about HOHH specifications. 

The rest of this thesis is organized as follows:  Chapter 2 presents
the HOHH specification logic and give examples of how this language can be
used to encode rule-based systems.  Chapter 3 provides the background
calculations necessary to generate strengthening lemmas correctly.
Chapter 4 describes the Abella Proof Assistant and its use, as well as
giving an example of proving a property of a specification written in the
specification language.  We finish with Chapter 5, which describes the
automatic generation of strengthening lemmas and the supporting
theorems, as well as the aforementioned new ``tactic command''.

%% file: hohh.tex
\chapter{Formalizing Relational Specifications}\label{ch:hohh}

In this chapter we describe the logic of higher-order hereditary
Harrop formulas (HOHH), the logic in whose context we will consider
automatically proving strengthening lemmas. The interest in this logic
arises from the fact that it is well-suited to formalizing and
prototyping software systems that are described in a rule-based
fashion. Indeed, HOHH provides the basis for the $\lambda$Prolog
programming language \cite{miller12proghol} that has been implemented,
for example, in the Teyjus system \cite{teyjus.website} and has been
used for exactly these purposes in many applications.
The strengthening lemmas that we want to prove turn out to be
useful in reasoning about the $\lambda$Prolog programs that result
from this process, an activity
that can be carried out using the Abella Proof Assistant.

The first section below presents the HOHH logic through its formulas
and its proof relation; the syntax of this logic is based on the
simply-typed $\lambda$-calculus that we digress briefly to also
describe. We then motivate the usefulness of HOHH in encoding
rule-based relational specifications. In preparation for a description
of the real technical content of this thesis, the concluding section
of the chapter explains what is meant by a strengthening lemma in the
context of the HOHH logic.

\section{Higher-Order Hereditary Harrop Formulas} \label{sec:HOHH}

In this section we describe the syntax and the inference rules that
define the logic of higher-order hereditary Harrop formulas and we
also discuss some metatheoretic properties of this logic that we will
find use for later. We use a somewhat simplified syntax for the
formulas in the logic from what is supported in the $\lambda$Prolog
language. We do this to simplify the exposition and we note that
nothing essential to the discussion is lost in the process.

\subsection{The Underlying Language}

The syntax of HOHH is based on the simply-typed $\lambda$-calculus
\cite{stlcChurch}. Two categories of expressions define the language:
types and terms. The types are built from atomic types, which
include built-in types and user-defined atomic types.  For example, to
work with binary trees containing natural numbers, a user might define
types $\mathsf{nat}$ to represent natural numbers and $\mathsf{bt}$ to
represent binary trees.  Higher-order types may be built using the
$\rightarrow$ type constructor, which takes two types and creates a
new type.  For example, $\sigma \rightarrow \tau$ is the type of a
function whose domain is the type $\sigma$ and whose co-domain is the
type $\tau$.  The $\rightarrow$ constructor is right-associative, so
the type $(\sigma_1 \rightarrow (\sigma_2 \rightarrow (... \rightarrow
(\sigma_n \rightarrow \tau)...)))$ can be written as $\sigma_1
\rightarrow \sigma_2 \rightarrow ... \rightarrow \sigma_n \rightarrow
\tau$.

The basis for the terms of the simply-typed $\lambda$-calculus
(\textit{$\lambda$-terms} or simply \textit{terms}) are a
countably-infinite set of variable symbols $\mathcal{V}$ and a
countably-infinite set of constant symbols $\mathcal{C}$.  Each member
of these sets is identified with a type. A member of
either of these sets is a $\lambda$-term by itself.  We can then build
larger terms using abstraction and application.  An abstraction of the
variable $x$ over a term $t$, written $\abs{x}{t}$, represents a
function where a term  may be given as an argument to produce a new
term.  An application of a term $t_1$ to a term $t_2$, written
$(\appTwo{t_1}{t_2})$, represents function application.  Repeated
application is left-associative, and so we can write
$(...(\appTwo{t_1}{t_2})...t_n)$ as $\appThree{t_1}{t_2}{...t_n}$.

Not all the terms that can be constructed in the manner described
above are considered well-formed. To be deemed well-formed, it must
also be possible to assign a type to the term. The rules for
determining the types can be seen in Figure \ref{fig:stlcTypingRules}.
We assume here that $\Sigma$ is a context that indicates the assigned
types for constants and variables.  From these we get the rule for
typing constants and variables, which is simply that they must be
identified in $\Sigma$ and that they then have the type assigned to
them.  To type an abstraction $\abs{x}{t}$ with the type $\sigma
\rightarrow \tau$ for some types $\sigma$ and $\tau$, the variable $x$
must have type $\sigma$ and $t$ must have the type $\tau$.  To show
that an application $(\appTwo{t_1}{t_2})$ has some type $\tau$, we
show that $t_1$ has the type $\sigma \rightarrow \tau$ for some type
$\sigma$ and that $t_2$ is of type $\sigma$.

\begin{figure}
  \begin{tabular}{m{1.5in} m{2.5in} m{2in}}
    \begin{prooftree}
      \AxiomC{$c : \tau \in \Sigma$}
      \RightLabel{const\_var}
      \UnaryInfC{$\Sigma \vdash c : \tau$}
    \end{prooftree}
&
    \begin{prooftree}
      \AxiomC{$\Sigma \vdash t_1 : \sigma \rightarrow \tau$}
      \AxiomC{$\Sigma \vdash t_2 : \sigma$}
      \RightLabel{app}
      \BinaryInfC{$\Sigma \vdash (\appTwo{t_1}{t_2}) : \tau$}
    \end{prooftree}
&
    \begin{prooftree}
      \AxiomC{$x : \sigma \in \Sigma$}
      \AxiomC{$\Sigma \vdash t : \tau$}
      \RightLabel{abs}
      \BinaryInfC{$\Sigma \vdash \abs{x}{t} : \sigma \rightarrow \tau$}
    \end{prooftree}
  \end{tabular}

  \caption{The typing rules for the simply-typed $\lambda$-calculus,
    where $\Sigma$ is a context assigning types to constants and variables, and $\vdash$
    represents the derivation of a type from the context $\Sigma$.}
  \label{fig:stlcTypingRules}
\end{figure}

In the discussions below we will need a substitution operation on
terms. If $x$ is a variable, $t_2$ is a term of the same type as $x$,
and $t_1$ is a term, we will denote the substitution of $t_2$ for $x$
in $t_1$ by $t_1[t_2/x]$. Since abstractions in the terms represent a
binding operation, we have to be careful about how such a substitution
is defined. One requirement is that we must not substitute for bound
variables: specifically, if $t_1$ is $\abs{x}{t_1'}$, then
$t_1[t_2/x]$ must be $t_1$, i.e. the substitution must leave the term
unchanged. Another requirement is that free variables in the term
being substituted must not end up being captured by an abstraction in the
term being substituted into: thus, if $t_1$ is $\abs{y}{t_1'}$ for
some $y$ different from $x$ and $y$ appears free in $t_2$, then we
must change $t_1$ to use a name that is different from $y$ and that
does not appear in $t_2$ and only then proceed to substituting $t_2$
in the body of the abstraction. We will not present this substitution
operation in detail here, assuming instead that the reader can
construct a definition for it based on our description of the main
difficulties that have to be accounted for.

Equality of terms is important in the logic used.  The three important
pieces of the equality relationship between terms are
$\beta$-conversion, $\eta$-conversion, and $\alpha$-conversion.
$\beta$-conversion involves terms with subterms of the form
$(\appTwo{(\abs{x}{t_1})} {t_2})$, called a $\beta$-redex.  The
$\beta$-redex can be replaced with $t_1[t_2/x]$ in
$\beta$-contraction.  Conversely, a term can be expanded to this form
in $\beta$-expansion.  $\eta$-conversion involves terms with subterms
of the form $\abs{x}{(\appTwo{t}{x})}$, where $x$ does not occur
free in $t$.  This is called an $\eta$-redex.  $\eta$-contraction is
replacing this term with $t$, while $\eta$-expansion is replacing a
term with a function type by an $\eta$-redex by wrapping it in an
abstraction.  The final piece is $\alpha$-conversion.  Terms include
names for bound variables, but these names do not matter in
themselves; the terms $\abs{x}{x}$ and $\abs{y}{y}$ both refer to
the identity function, but are not exactly the same by variable names
and so are not seen as equal.  To show their equality,
$\alpha$-conversion is used to rename variables.  For the terms
$\abs{x}{t_1}$ and $\abs{y}{t_2}$, a new variable $z$ is chosen that is
free in both $t_1$ and $t_2$.  The original terms are then replaced by
$\abs{z}{t_1[z/x]}$ and $\abs{z}{t_2[z/y]}$.  By renaming
variables in this way, two terms with the same structure that had
different variable names can be seen to be the same.

It is possible to $\beta$-contract and $\eta$-contract a term to a
point where these cannot be applied anymore, which is called the
$\beta\eta$-normal form of a term.  Two terms can be contracted to
their unique $\beta\eta$-normal forms, then $\alpha$-conversion can be
used to give them the same names for bound variables.  If they are
equal after this, the two original terms are equivalent.

To build a logic based on the simply-typed $\lambda$-calculus, we
first identify a special atomic type $\mathsf{o}$ that will function
as the type of logical formulas. We  then add several {\it logical}
constants: $\Rightarrow$: $\mathsf{o} \rightarrow \mathsf{o}
\rightarrow \mathsf{o}$, representing implication; \&: $\mathsf{o}
\rightarrow \mathsf{o} \rightarrow \mathsf{o}$, representing
conjunction; $\top$: $\mathsf{o}$, representing truth; $\bot$: $\mathsf{o}$,
representing falsity; and $\Pi_{\tau}$: $(\tau \rightarrow \mathsf{o})
\rightarrow \mathsf{o}$, representing universal quantification over
the type $\tau$.  The constant $\Pi$ stands for an infinite set of
constants, with a different one for each type $\tau$.  We will usually
not include the type subscript when writing quantifications.  Logical
formulas are then built using these constants and user-defined
predicates.

We can construct a logic over the language we have described by
introducing rules corresponding to each of the logical constants that
allow us to construct proofs for formulas. In a fully configured
logic, we would not place any restrictions on the forms of the
formulas we want to derive. The HOHH logic takes a different view: it
limits the kinds of formulas permitted with a goal of allowing
specialized inference rules that are motivated by the desire to
construct derivations that parallel those in rule-based systems. More
specifically, this logic is concerned with two classes of formulas
that are called {\it goal formulas} and {\it program clauses}. These
formulas are described by the following syntax rules in which goal
formulas and program clauses are denoted by $G$ and $D$ respectively:
\begin{center}
  $G ::= \top \ | \ A_r \ | \ G \ \& \ G \ | \ D \Rightarrow G \ | \ \Pi_{\tau} x. G$\\
  $D ::= G \Rightarrow A_r \ | \ \Pi_{\tau} x. D$
\end{center}
In these rules, we take $A_r$ to represent a formula whose leftmost
non-parenthesis symbol is a predicate constant different from the
logical constants. A formula of this kind is also referred to as a
{\it rigid atom}. To write multiple universal
quantifications for variables $x_1, x_2, ..., x_n$ with corresponding
types $\tau_1, \tau_2, ..., \tau_n$, we write $\Pi \bar{x} :
\bar{\tau}$.  

In the context of the HOHH logic, we think of a collection of program
clauses as a \emph{specification} or \emph{program}. 
In later discussions, we will need to refer to the atom $A$ in a
formula $F$ of the form $\Pi \bar{x} : \bar{\tau}. G \Rightarrow A$
as  the head of $F$, written $\mathcal{H}(F)$.  Further, we will call the predicate
head of $A$, i.e. the leftmost non-parenthesis symbol in $A$, the head
predicate of $F$ as well and we will use the notation
$\headPred{F}$ to refer to it.  Finally, we have the body of $G$,
a goal formula, written $\mathcal{L}(G)$.  If $G$ is an implication
$D \Rightarrow G^{\prime}$, $\mathcal{L}(G) = \{D\}$; otherwise,
$\mathcal{L}(G) = \emptyset$.

\subsection{The Specification Logic}

\begin{figure}
  \fontsize{10pt}{10pt}
    \begin{tabular}{m{1.1in} | m{2.5in} m{2.5in}}
      Goal-Reduction
&
      \begin{prooftree}
        \AxiomC{}
        \RightLabel{$\top R$}
        \UnaryInfC{$\Sigma; \Gamma; \Delta \vdash \top$}
      \end{prooftree}
&
      \begin{prooftree}
        \AxiomC{$\Sigma; \Gamma; \Delta \vdash G_1$}
        \AxiomC{$\Sigma; \Gamma; \Delta \vdash G_2$}
        \RightLabel{$\& R$}
        \BinaryInfC{$\Sigma; \Gamma; \Delta \vdash G_1 \& G_2$}
      \end{prooftree}
\\ &
      \begin{prooftree}
        \AxiomC{$\Sigma; \Gamma; \Delta, D \vdash G$}
        \RightLabel{$\Rightarrow\!R$}
        \UnaryInfC{$\Sigma; \Gamma; \Delta \vdash D \Rightarrow G$}
      \end{prooftree}
&
      \begin{prooftree}
        \AxiomC{$c \notin \Sigma$}
        \AxiomC{$\Sigma, c : \tau; \Gamma; \Delta \vdash B[c/x]$}
        \RightLabel{$\Pi R$}
        \BinaryInfC{$\Sigma; \Gamma; \Delta \vdash \Pi_{\tau}x B$}
      \end{prooftree}
\\ \hline
      Backchaining
&
      \begin{prooftree}
        \AxiomC{$\Sigma; \Gamma; \Delta \vdash G$}
        \AxiomC{$\Sigma; \Gamma; \Delta; [A] \vdash A$}
        \RightLabel{$\Rightarrow\!L$}
        \BinaryInfC{$\Sigma; \Gamma; \Delta; [G \Rightarrow A] \vdash A$}
      \end{prooftree}
&
      \begin{prooftree}
        \AxiomC{$\Sigma \vdash t : \tau$}
        \AxiomC{$\Sigma; \Gamma; \Delta; [F\ t] \vdash A$}
        \RightLabel{$\Pi L$}
        \BinaryInfC{$\Sigma; \Gamma; \Delta; [\Pi_{\tau} F] \vdash A$}
      \end{prooftree}
\\ \hline
      Structural
&
      \begin{prooftree}
        \AxiomC{}
        \RightLabel{\init}
        \UnaryInfC{$\Sigma; \Gamma; \Delta; [A] \vdash A$}
      \end{prooftree}
&
      \begin{prooftree}
        \AxiomC{$D \in \Gamma \cup \Delta$}
        \AxiomC{$\Sigma; \Gamma; \Delta; [D] \vdash A$}
        \RightLabel{\focus}
        \BinaryInfC{$\Sigma; \Gamma; \Delta \vdash A$}
      \end{prooftree}
    \end{tabular}

  \caption{The inference rules of HOHH, where $\Sigma$ contains type assignments and $\Gamma$ and $\Delta$ are sets of clauses to be used in proving the goal.  In the $\&R$ rule, $i$ is either 1 or 2.}
  \label{fig:HOHHinfRules}
\end{figure}

In Figure \ref{fig:HOHHinfRules}, we see the inference rules of
HOHH. These rules are oriented around proving judgments of the form
$\Sigma; \Gamma; \Delta \vdash G$ that are called \emph{sequents}. In
such a sequent, $\Sigma$ is a \emph{signature} that contains the
constants that appear in the sequent with each constant being paired
with its type, $\Gamma$ is a set of program clauses that is referred
to as the \emph{static context} of the sequent, and $\Delta$ is another set
of program clauses that is referred to as the
\emph{dynamic context} of the sequent. Intuitively, such a sequent asserts
that the formula $G$, also called the goal formula of the
sequent, holds whenever the program clauses in $\Gamma$ and $\Delta$
hold. Initially, $\Delta$ is an empty set, $G$ is the formula we want
to show holds, and $\Gamma$ is a specification in whose context we want
to show $G$ holds; $\Sigma$ is set assigning types to the collection of constants
appearing in $\Gamma$ and $G$. When searching for a proof of such a
sequent, we first simplify the goal formula using the pertinent
\emph{goal-reduction} rule. In the course of using these rules, we may
add new constants to $\Sigma$ through applications of the $\Pi R$ rule
and new formulas to the dynamic context through applications of the
$\Rightarrow\!R$ rule. When this formula has been reduced to an
atomic one, the search switches to a \emph{backchaining} mode. This
begins with our first selecting a program clause from $\Gamma$ or
$\Delta$ using the \focus\ rule. This rule introduces a sequent of the
form $\Sigma; \Gamma; \Delta; [D] \vdash A$ that has the same
assertional content as our regular sequents with the exception that it
also signals our intent to work with a particular program
clause---called the focus---in
looking for a proof. The backchaining rules are used to further this
intent. The $\Rightarrow\!L$ rule may spawn an attempt to prove the
goal formula that constitutes the ``body'' of the chosen clause and 
success in using the program clause depends on our being able to match
the head of an instance of the program clause with the atomic goal formula
using the \init\ rule. 


\subsection{Metatheoretic Properties of the Logic} \label{sec:metatheoretic}

It has been shown that the derivation system of HOHH corresponds to
provability in intuitionistic logic \cite{uniformProofs}. Thus, any
sequent of the kind we are considering has a derivation in this system
if and only if it is valid in intuitionistic logic.  Then any
metatheoretic properties of intuitionistic logic, when limited to
considering sequents with only program clauses and goal formulas in
the relevant places, also apply to this derivation system, and we can
think of using them in constructing HOHH derivations.  

One such property is instantiation, which is if $\Sigma \vdash t :
\tau$ and $\Sigma, c : \tau; \Gamma; \Delta \vdash G$ where $c$ is not
free in $\Gamma$ are derivable,  then it is also possible to derive
$\Sigma; \Gamma; \Delta[t/c] \vdash G[t/c]$, where we allow $[t/c]$ to
be the capture-avoiding substitution of $t$ for $c$ in a formula or
set of formulas.  More simply, if $t$ has type $\tau$, then a
derivation that includes a constant of this type that is not found in
the static context is also valid if $t$ is used in the constant's
place. 

The monotonicity property of intuitionistic derivability states that
if $\Sigma; \Gamma; \Delta \vdash G$ is derivable and $F \in \Delta$
implies $F \in \Theta$, then $\Sigma; \Gamma; \Theta \vdash G$ also
derivable. From this, it follows that if we have succeeded in proving
a sequent with multiple copies of a formula in the dynamic context,
then a sequent in which we remove one of the copies will still be
derivable. This metatheoretic property is a special case of a more
general property known as the admissibility
of contraction. Similarly, it follows that adding formulas to the
dynamic context of a derivable sequent yields another sequent that is
derivable. This property is a special case of a property known as the
admissibility of weakening. 

\section{Encoding Rule-Based Systems} \label{sec:encodingRules}

Many notions that are of interest in a computational setting can be
described via relations that are presented in a rule-based fashion. As
an example, consider the task of appending two lists to produce a
third. We can capture the intent of this computation through a
relation between three lists. Moreover, we can describe this relation
completely by saying that it holds if and only if it can be derived
using the following rules:
\begin{center}
  \begin{tabular}{m{2in}m{3in}}
    \begin{prooftree}
      \AxiomC{}
      \RightLabel{app-nil}
      \UnaryInfC{$\appFour{append} {nil} {L} {L}$}
    \end{prooftree}
    &
    \begin{prooftree}
      \AxiomC{$\appFour{append} {L_1} {L_2} {L_3}$}
      \RightLabel{app-cons}
      \UnaryInfC{$\appFour{append} {(X::L_1)} {L_2} {(X::L_3)}$}
    \end{prooftree} \\
  \end{tabular}
\end{center}
To understand that these rules are a complete description of the
relation, we note that this is a definition that is inductive on the
structure of the first argument of {\it append}. 
Note also that in the usual interpretation, we intend rules such as
these to be interpreted not only as a means for deriving the {\it
  append} relation but also as the {\it only} means for doing so.

These rules can be encoded in HOHH using program clauses.  We will
assume an encoding of the natural numbers in a type $\mathsf{nat}$,
which will be the type of elements in the lists being appended.  We
will also assume an encoding of type $\mathsf{list}$, with
constructors $\mathsf{nil} : \mathsf{list}$ and $\mathsf{cons} :
\mathsf{nat} \rightarrow \mathsf{list} \rightarrow \mathsf{list}$,
with $\mathsf{nil}$ representing the empty list and $\mathsf{cons}$
representing list construction.  A predicate symbol $\mathsf{append} :
\mathsf{list} \rightarrow \mathsf{list} \rightarrow \mathsf{list} \rightarrow \mathsf{o}$ is
created to represent the append relationship.  Then the rules as shown
above can be translated to program clauses as follows: 
\[\appFour{\mathsf{append}} {\mathsf{nil}} {L} {L}
        \hspace*{15mm}
  \appFour{\mathsf{append}} {L_1} {L_2} {L_3} \Rightarrow \appFour{\mathsf{append}} {(\appThree{\mathsf{cons}}{X}{L_1})} {L_2} {(\appThree{\mathsf{cons}} {X} {L_3})}\]
We use the logic programming convention that the variables that start
with capital letters are universally quantified over the whole
formula. 

Using these two program clauses, we can determine whether the append
relation holds between three lists $l_1$, $l_2$, and $l_3$.  To do so,
the three lists are encoded into the $\mathsf{list}$ type, then we try to
prove the sequent $\Sigma; \Gamma; \emptyset \vdash
\appFour{\mathsf{append}} {l_1} {l_2} {l_3}$, where $\Sigma$ contains
the constants for building lists and the $\mathsf{append}$ predicate,
and $\Gamma$ contains the program clauses containing the rules for
$\mathsf{append}$.  Clearly the rules and program clauses are
equivalent, and carrying out this derivation would also follow the
same structure as using the rules.  Then the behavior of the
rule-based $append$ is the same as is captured in the program clauses
for $\mathsf{append}$.

Another example of a rule-based specification is assigning types to
terms in the simply-typed $\lambda$-calculus.  The typing relationship
to be defined is $\Gamma \vdash t : \tau$, where $t$ is a term of type
$\tau$ and $\Gamma$ is a typing context that has the form $x_1 :
\tau_1, ..., x_n : \tau_n$ for distinct variables $x_1, ..., x_n$ of
types $\tau_1, ..., \tau_n$.  The typing rules are as follows: 

\begin{center}
  \begin{tabular}{m{1.5in} m{2.5in} m{2in}}
    \begin{prooftree}
      \AxiomC{$x : \tau \in \Gamma$}
      \RightLabel{varTy}
      \UnaryInfC{$\Gamma \vdash x : \tau$}
    \end{prooftree}
&
    \begin{prooftree}
      \AxiomC{$\Gamma \vdash t_1 : \sigma \rightarrow \tau$}
      \AxiomC{$\Gamma \vdash t_2 : \sigma$}
      \RightLabel{appTy}
      \BinaryInfC{$\Gamma \vdash (\appTwo{t_1}{t_2}) : \tau$}
    \end{prooftree}
&
    \begin{prooftree}
      \AxiomC{$\Gamma, x : \sigma \vdash t : \tau$}
      \RightLabel{absTy}
      \UnaryInfC{$\Gamma \vdash \abs{x}{t} : \sigma \rightarrow \tau$}
    \end{prooftree}
  \end{tabular}
\end{center}

We create an encoding of types and terms, with the encoding of types having type
$\mathsf{ty}$ and the encoding of terms having type $\mathsf{tm}$.
Atomic types are represented by $\mathsf{b} : \mathsf{ty}$, and we
have the arrow type constructor $\mathsf{arr} : \mathsf{ty}
\rightarrow \mathsf{ty} \rightarrow \mathsf{ty}$ to represent function
types.  Terms are encoded by $\mathsf{app} : \mathsf{tm} \rightarrow
\mathsf{tm} \rightarrow \mathsf{tm}$ for application and $\mathsf{abs}
: \mathsf{ty} \rightarrow (\mathsf{tm} \rightarrow \mathsf{tm})
\rightarrow \mathsf{tm}$ for abstraction.  In the abstraction
encoding, we allow the abstraction available in the underlying logic
used for encoding to handle the binding for us.  Doing this allows us
to allow the underlying logic to handle the scoping of bindings and
the substitution of terms in an object-language term.  As an example
of the encoding, the term $(\abs{(x:b)}{\abs{(y:(b \rightarrow
    b))}{\appTwo{y}{x}}})$ is encoded as $\appThree{\mathsf{abs}}
{\mathsf{b}}  {(\abs{x}{\appThree{\mathsf{abs}}
    {(\appThree{\mathsf{arr}}{\mathsf{b}}{\mathsf{b}})}
    {(\abs{y}{\appThree{\mathsf{app}}{y}{x}})}})}$. 

We create a predicate $\mathsf{typeof} : \mathsf{tm} \rightarrow
\mathsf{ty} \rightarrow \mathsf{o}$ to represent the typing relation.
The typing context will be held in the dynamic context in derivations,
so it need not be included in the predicate.  Then the typing rules
are 

\begin{center}
  \begin{tabular}{c}
    $\appThree{\mathsf{typeof}} {M_1} {(\appThree{\mathsf{arr}}{T_1}{T_2})} \Rightarrow \appThree{\mathsf{typeof}}{M_2}{T_1} \Rightarrow \appThree{\mathsf{typeof}}{(\appThree{\mathsf{app}}{M_1}{M_2})}{T_2}$ \\
    $(\Pi x. (\appThree{\mathsf{typeof}} {x} {T_1} \Rightarrow \appThree{\mathsf{typeof}} {(\appTwo{M}{x})} {T_2})) \Rightarrow \appThree{\mathsf{typeof}} {(\appThree{\mathsf{abs}} {T_1} {M})} {(\appThree{\mathsf{arr}} {T_1} {T_2})}$
  \end{tabular}
\end{center}

We assume that any variables starting with capital letters are
implicitly universally quantified over the whole formula.  Then
showing that the typing relation $\Gamma \vdash M : \tau$ holds, we
show that the sequent $\Sigma; \Gamma; \emptyset \vdash
\appThree{\mathsf{typeof}} {\bar{M}} {\bar{\tau}}$ is derivable, where
$\Sigma$ contains the constants for creating terms and types, along
with the $\mathsf{typeof}$ predicate, $\Gamma$ contains the two
program clauses defining the $\mathsf{typeof}$ predicate, and
$\bar{M}$ and $\bar{\tau}$ represent the encodings of $M$ and $\tau$. 

Consider the derivation of $\appThree{\mathsf{typeof}}
{(\appThree{\mathsf{abs}}{\mathsf{b}}{\abs{x}{x}})}
{(\appThree{\mathsf{arr}}{\mathsf{b}}{\mathsf{b}})}$.  We start with
an atomic goal, so the goal-reduction rules do not need to be
used. Then the derivation is started by focusing on the program clause
for typing abstractions.  After instantiating the
universally-quantified variables to match the goal, we see that the
formula we are focusing on is 
\[\Pi x.(\appThree{\mathsf{typeof}}{x}{\mathsf{b}} \Rightarrow \appThree{\mathsf{typeof}}{x}{\mathsf{b}}) \Rightarrow \appThree{\mathsf{typeof}} {(\appThree{\mathsf{abs}}{\mathsf{b}}{\abs{x}{x}})} {(\appThree{\mathsf{arr}}{\mathsf{b}}{\mathsf{b}})}\]
We then use the $\Rightarrow\!L$ rule to reduce the focused formula to
atomic form.  This splits the derivation, as we have both to prove the
current goal focused on the head formula and the antecedent.  The
current goal matches the head formula of the focused formula, and is
then proved immediately by the \init~rule. 

Our other derivation to be completed is that of
\[\Sigma; \Gamma; \emptyset   \vdash   \Pi x.(\appThree{\mathsf{typeof}}{x}{\mathsf{b}} \Rightarrow \appThree{\mathsf{typeof}}{x}{\mathsf{b}})\]
The $\Pi R$ rule is used to introduce a new constant that is added to
the signature.  Since the top-level logical connective on the right is
now an implication, the $\Rightarrow\!R$ rule is used to reduce it to
its head, giving us the sequent 
\[\Sigma, x:\mathsf{tm}; \Gamma; \appThree{\mathsf{typeof}}{x}{\mathsf{b}}   \vdash   \appThree{\mathsf{typeof}}{x}{\mathsf{b}}\]
In this way $\appThree{\mathsf{typeof}} {x} {\mathsf{b}}$ is added to
the dynamic context, saving the assumption of the type of the variable
$x$ for use in the remainder of the derivation.  This obviates
tracking these assumptions in the $\mathsf{typeof}$ predicate itself.
By focusing on the formula in the dynamic context, we are able to use
the \init~rule, finishing this branch of the derivation as well.  Since
both branches are completed, the whole derivation is completed, and
the original goal has been proven. 

\section{Strengthening Lemmas}

Our interest in this thesis is in strengthening lemmas in the context
of the HOHH logic. These lemmas take the following form: Suppose we
know that $\Sigma;\Gamma; \Delta, F \vdash G$ is derivable. Moreover,
suppose that we can determine that $F$ could not possibly have been
used in this derivation. Then we can conclude that the sequent 
$\Sigma; \Gamma; \Delta \vdash G$ must also have a derivation.

A critical part of the reasoning described above is showing that the
assumption $F$ that we want to ``discard'' could not figure in the
derivation of $G$. To do this, we find all the possible forms of goals
which may arise in proving $G$, and all formulas that may occur in the
context while proving $G$, and show that $F$ cannot be used for any of
the goals or by using any of the formulas that may occur in the
context.  The methods for finding all formulas that may be in the
context and all possible goals are described in the next chapter. Here
we limit ourselves to discussing some of the issues that must be
considered in designing such a method.

As a small example, consider the sequent $\Sigma; \emptyset; F
\Rightarrow G, F \vdash G$.  Clearly, by focusing and backchaining on
the first formula in the context, the sequent becomes $\Sigma;
\emptyset; F \Rightarrow G, F \vdash F$, so in this case $G$ cannot
be strengthened from $F$.  Another, more subtle, example of when
strengthening fails is the sequent $\Sigma; \emptyset; F \Rightarrow
A, A \Rightarrow G, F \vdash G$.  By focusing and backchaining on
$A \Rightarrow G$, the goal formula becomes $A$, from which it is
possible to focus and backchain on $F \Rightarrow A$, once again
giving a goal of $F$, which can be solved by the $F$ in the context.
Then it should be clear why considering all goals that may arise in the
course of computation, rather than just the original goal, is
important. 

If we consider the sequent $\Sigma; \emptyset; F \Rightarrow B, (B
\Rightarrow A) \Rightarrow G, A, F \vdash G$, we can see that
strengthening from $F$ is possible.  The only action available to
start is to focus and backchain on $(B \Rightarrow A) \Rightarrow G$,
which gives a goal of $B \Rightarrow A$.  Using the $\Rightarrow\!R$
rule, we get the goal $A$ and the formula $B$ is added to the context.
This is then solved by focusing and backchaining on its instance in
the context.  It can be seen that $F$ can never become a goal in the
derivation of the original sequent, since that would require
backchaining with a goal of $B$.  Since $F$ can never become a goal,
we are able to strengthen from $F$ to get the sequent $\Sigma;
\emptyset; F \Rightarrow B, (B \Rightarrow A) \Rightarrow G, A \vdash
G$.  Similarly, since $B$ can never become a goal, it is also possible
to strengthen from $F \Rightarrow B$, giving the sequent $\Sigma;
\emptyset; (B \Rightarrow A) \Rightarrow G, A \vdash G$. 

An example of the use of strengthening is type independence.  If we
have two types, $\tau_1$ and $\tau_2$, then $\tau_2$ is independent of
$\tau_1$ if, whenever the typing judgment $\Sigma, x : \tau_1 \vdash t
: \tau_2$ holds, $\Sigma \vdash t : \tau_2$ also holds.  Type
independence is less conservative than type subordination, as
discussed in \cite{yutingPaper}, so we are able to more closely
approximate which types a given type depends on, allowing us to prune
unnecessary dependencies. 



%% file: dependencies.tex
\chapter{Calculating Predicate Dependencies}\label{ch:dependencies}

Our ultimate goal is to show that some assumptions are unnecessary in
proving a given goal.  To do this, we must determine what formulas may
arise in the dynamic context during a proof, along with determining
what types of goals may arise in the course of any derivation. In this
chapter, we present the algorithms developed in \cite{yutingPaper} for
computing both these sets. Using the results of these algorithms, we
are able to determine when a strengthening lemma holds. Moreover, this
information is also useful in developing an explicit proof of the
strengthening lemma, a task that we take up in
Chapter~\ref{ch:strengthening-proof}. 

\section{Calculating Dynamic Contexts} \label{sec:dynCtx}

We begin by considering the dynamic context of a predicate.  If we
have an implication as our goal, then the $\Rightarrow\!R$ rule adds
the antecedents of this goal to the context while reducing the goal
formula to the head of this formula.  These antecedents are then
available to be used in proving the new goal, or any goal that further
arises in the derivation.  Using the $\Rightarrow\!L$ backchaining
rule, we can get goal formulas with different head predicates.  The
new goal then has the same context as the previous goal had.  In this
way, we find that the dynamic context of one predicate may contain the
dynamic context of another predicate.   

\begin{figure}
  \begin{center}
    \begin{minipage}{5.5in}
      \begin{tabbing}
        \qquad \= \kill
          Let $\Gamma'$ be a finite set equal to $\Gamma$ and
              $\mathcal{C} \leftarrow \emptyset$\\
          \textbf{while $\Gamma' \neq \emptyset$ do}\\
        \>   pick some $D = (\Pi \bar{x}. (G_1\ \&\ ...\ \&\ G_n)
                \Rightarrow A)$ from $\Gamma'$\\ 
        \>   add equations $\{ C(\mathcal{H}_p(G_i)) =
                C(\mathcal{H}_p(G_i)) \cup C(\mathcal{H}_p(A)) \cup
                \mathcal{L}(G_i)\ |\ i = 1..n\}$ to $\mathcal{C}$\\ 
        \>   remove $D$ from $\Gamma'$ and add clauses in $\bigcup_{i
                \in 1..n} \mathcal{L}(G_i)$ to $\Gamma'$\\
          \textbf{end while}
      \end{tabbing}
    \end{minipage}
  \end{center}
  \vspace*{-5mm} 

  \caption{Algorithm for collecting constraints on dynamic contexts}
  \label{fig:ctxAlg}
\end{figure}

In Figure \ref{fig:ctxAlg} we see an algorithm for calculating
constraints on the dynamic context of a predicate.  We assume $\Gamma$
is a finite set of program clauses.  We let $C(a)$ represent a set of
formulas that can occur in the dynamic context of the predicate $a$,
that is, formulas that may be in the dynamic context when the goal is
$G$ with $\headPred{G} = a$, and $\mathcal{C}$ represent the set of
all constraint equations for all predicates.  The only way we add
formulas to the context is through the $\Rightarrow\!R$ rule, and this
can happen with any formula currently in the context with a head that
matches the current goal's head.  To compute the set of constraints,
we go through every formula from $\Gamma$ and their subformulas.  For
each formula $D = (\Pi \bar{x}. (G_1\ \&\ ...\ \&\ G_n) \Rightarrow
A)$, we add a new constraint equation to $\mathcal{C}$ for
$\headPred{G_i}$ consisting of the union of $C(\headPred{G_i})$,
$C(\headPred{A})$, and $\mathcal{L}(G_i)$ for $i = 1..n$.  We include
the dynamic context of $\headPred{A}$ because, when backchaining on
$D$, the current goal's head predicate must be $\headPred{A}$, and so
the formulas in the dynamic context of $\headPred{A}$ can also be in
the dynamic context for the derivations of $G_i$ for $i = 1..n$.  We
include $\mathcal{L}(G_i)$ because, in reducing the goal $G_i$ to
atomic form, all the formulas of the body will be moved into the
dynamic context by the $\Rightarrow\!R$ rule.  Since we have accounted
for this formula, we remove it from further consideration and add in
the formulas from the bodies of all the antecedents to calculate their
effects on the dynamic contexts.

Once we have finished collecting the constraint equations, we need to
iterate over the constraint equations to find the full set of formulas
that may occur in the dynamic context of each predicate.  We start
with $C(a) = \emptyset$ for all $a$, where $a$ is a predicate that
occurs in $\Gamma$.  We repeatedly apply each of the constraint
equations iteratively until no new formulas are added to any dynamic
context.  At this point $C(a)$ will be a set containing all the
formulas that the dynamic context may contain for each predicate $a$.

\section{Calculating Predicate Dependencies} \label{sec:dependencies}  

Once we know what formulas can occur in the dynamic contexts of
predicates, we can determine the dependencies between different
predicates.  To find the dependencies of a predicate $a$, we look at
the formulas that occur in both its dynamic context and the program
clauses, since these are the formulas it may focus and backchain on in
the course of a derivation.  If a formula
$D = (\Pi \bar{x}. (G_1\ \&\ ...\ \&\ G_n) \Rightarrow A)$ is used to backchain,
then the provability of the current goal, the head predicate of which
is $a$, depends on the provability of the head predicate of $G_i$ for $i = 1..n$.
Thus $a$ depends on $\headPred{G_i}$, which may also
depend on other predicates.  Then, since the provability of $a$
depends on the provability of $\headPred{G_i}$, the provability of
which depends on a set of other predicates, $a$ depends on these other
predicates as well. 

\begin{figure}
  \begin{center}
    \begin{minipage}{5.75in}
      \begin{tabbing}
        \qquad \= \kill
          Let $\mathcal{S} \leftarrow \emptyset$\\
          \textbf{for all} $a \in \Delta$ \textbf{do}\\
        \>   \textbf{for all} $D \in \Gamma \cup C(a)$ where $D = (\Pi
                \bar{x}. (G_1\ \&\ ...\ \&\ G_n) \Rightarrow A)$ and
                $\headPred{A} = a$ \textbf{do}\\ 
        \>   add $(S(a) = S(a) \cup \bigcup_{i = 1..n}
                S(\headPred{G_i}))$ to $\mathcal{S}$\\
        \>   \textbf{end for}\\
          \textbf{end for}
      \end{tabbing}
    \end{minipage}
  \end{center}
  \vspace*{-5mm} 

  \caption{Algorithm for collecting constraints on dependencies}
  \label{fig:depAlg}
\end{figure}

The algorithm for computing constraints on dependencies is found in
Figure \ref{fig:depAlg}.  We once again assume $\Gamma$ is a set of
program clauses.  We let $\Delta$ be the set of all predicates that
occur in $\Gamma$, $\mathcal{S}$ be a set of equations constraining
the dependency relations, and $S(a)$ be the set of predicates $a$
depends on, where $a$ is a predicate.  Then, for each of these
predicates $p$, we iterate over the full context that $p$ may have,
both the dynamic context and the program clauses.  For each formula
$D$, if $\mathcal{H}_p(D) = p$, a goal with $p$ as its head predicate
could successfully backchain on it, and then all antecedents of $D$
would have to be proven, so we add a constraint equation for $p$ that
adds the dependencies of the head predicate for each antecedent to the
dependencies for $p$.

As in the case of calculating the dynamic contexts, we must iterate
over the constraint equations to find the full sets of dependencies.
We start with $S(a) = \{a\}$ for all $a \in \Delta$, since a predicate
must depend on itself.  We then iteratively apply the constraint
equations in $\mathcal{S}$ until no new predicates are added to any
dependency set.  Then we know that $S(a)$ contains all the
dependencies of $a$, and, in the course of proving a goal $G$ where
$\mathcal{H}_p(G) = a$, a goal cannot arise with a head predicate that
is not in $S(a)$.

\section{The Conservativity of Our Computations} \label{sec:compConserv}

The computations discussed in the first two sections of this chapter
capture all possible formulas that may occur in the dynamic context of
a predicate and all predicates it may depend on, but they may
overestimate these dependencies.  This occurs because the computations
don't take into account the fact that some of the formulas do not
occur in the same branch of computation, and yet they are used
together in calculating the dependencies. 

Let us consider an example consisting of the sole formula 
\[(((s \Rightarrow r) \Rightarrow p)\ \&\ ((r \Rightarrow p)
\Rightarrow p)) \Rightarrow q\]  
It is clear that if we are deriving $\Sigma; \Gamma; \Delta \vdash q$
and backchain on this formula, we will have two new goals to show,
$\Sigma; \Gamma; \Delta, s \Rightarrow r \vdash p$ and $\Sigma;
\Gamma; \Delta, r \Rightarrow p \vdash p$.  Then it is clear that $s
\Rightarrow r$ and $r \Rightarrow p$ both occur in the dynamic context
of $p$, but they cannot occur in the dynamic context at the same time.
However, in computing the dependencies, these branches are ignored and
both are considered together.  Then we get dependency constraint
equations $S(r) = S(r) \cup S(s)$ from $s \Rightarrow r$ and $S(p) =
S(p) \cup S(r)$ from $r \Rightarrow p$, as well as $S(q) = S(q) \cup
S(p) \cup S(p)$.  Simplifying these to get the full calculated
dependencies, we get the following:
\begin{center}
  $S(s) = \{s\}$, $S(r) = \{r, s\}$, $S(p) = \{p, r, s\}$,
           $S(q) = \{q, p, r, s\}$
\end{center}
\noindent If we consider the goals we can have by focusing and
backchaining in derivations starting with the goal $q$ as above,
however, we see that we have nothing to backchain on in ${\Sigma;
  \Gamma; \Delta, s \Rightarrow r \vdash p}$, and we can only get
$\Sigma; \Gamma; \Delta, r \Rightarrow p \vdash r$ from $\Sigma;
\Gamma; \Delta, r \Rightarrow p \vdash p$.  Thus we can see that the
goal $s$ cannot arise, and $p$, $q$, and $r$ do not actually depend on
$s$, but we include it in our calculated dependencies for them. 

The overestimation of dependencies is due to the inclusion together in
the dynamic context of formulas that would appear in different
branches of the computation.  It is important to note that computing a
larger set than will actually arise does not make our procedure for
determining when a strengthening lemma holds an unsound one: such a
lemma will certainly hold whenever the procedure says it does, an
observation that follows from the metatheoretic monotonicity property
discussed in Subsection \ref{sec:metatheoretic}.. Rather, 
what it does is it sometimes prevents us from providing a positive
answer when in fact a more targeted analysis would allow us to do so. 

We could develop a more precise algorithm for calculating dependencies
that overcomes the specific issue highlighted by the example in this
section. The way to deal with this is to keep track of which type of
branch is being taken to reach a certain goal and particularize the
sets to the relevant branches.  In the prior example,
we would need to track whether we were carrying out the proof for $((s
\Rightarrow r) \Rightarrow p)$ or $((r \Rightarrow p) \Rightarrow p)$.
Then we could avoid the issue of adding together contexts from
different branches, and thus also avoid the issue of overestimating
the dependencies. We note that keeping track of the separate branches
would complicate the calculations of the sets. More importantly, it would also
complicate the process of proving the strengthening lemma, requiring
many auxiliary strengthening lemmas that are indexed not only by
predicates but also by the branches in which we are considering their
derivation. We have not explored the description of a more precise
algorithm for calculating dependencies because we are not convinced at
this stage that the more complicated proof structure will be
compensated for by the ability to prove more strengthening lemmas
automatically in practice.

%% file: abella.tex
\chapter{The Abella Proof Assistant}

Our goal now is to use the information that is computed by the
algorithms described in the previous chapter to produce explicit proofs
of strengthening lemmas. We will construct these proofs within the
framework of the Abella Proof Assistant. The reason for our picking
this framework is twofold. First, Abella actually encodes the HOHH
logic and provides us a means for reasoning about derivations within
it; thus, it \emph{is} a framework within which we are able to carry out the
task that is of interest. Second, the strengthening lemmas that we
want to prove are often motivated by other proofs related to HOHH
specifications that we want to construct using the Abella system. This
was, in fact, one of the original motivations for considering these
strengthening lemmas.

In this chapter we introduce the Abella Proof Assistant towards setting
up a context for describing the automatic generation of proofs for
strengthening lemmas. We begin by describing the logic that underlies
Abella. We then explain the means Abella provides for constructing
proofs within this logic. In the last two sections, we present the
encoding of the HOHH logic within Abella and we explain how Abella
allows us to reason about derivability in the HOHH logic.


\section{The Logic Underlying Abella}

The language used by the logic that Abella implements is also based on
the simply typed $\lambda$-calculus. The types used are determined in
a similar fashion to that in the HOHH language. Like in HOHH, there is
a type for formulas with the difference that this type is named
$\mathsf{prop}$ rather than $\mathsf{o}$.  Once again, the language
contains a special collection of logical constants for constructing
formulas. Specifically, these are $\top$ and
$\bot$, both of type $\mathsf{prop}$; $\wedge$, $\vee$, and $\supset$
of type $\mathsf{prop} \rightarrow \mathsf{prop} \rightarrow
\mathsf{prop}$; $\forall_{\tau}$ and $\exists_{\tau}$, both of type
$(\tau \rightarrow \mathsf{prop}) \rightarrow \mathsf{prop}$; and
$=_{\tau}$, of type $\tau \rightarrow \tau \rightarrow \mathsf{prop}$.
The last three symbols, which represent universal quantification,
existential quantification, and equality, respectively, actually denote
infinite sets of constants, with a  different constant for each
type $\tau$.  We will generally drop the
type subscript when writing these symbols, assuming that their types
can be inferred from the context. In writing quantified formulas, we
will abbreviate $\forall\ (\lambda x.F)$ by $\forall x.F$ and similarly
$\exists\ (\lambda x. F)$ by $\exists x.F$. If we are quantifying
multiple variables $x_1, ..., x_n$, we will write them as $\bar{x}$.
For example, $\forall x_1 \forall x_2 ... \forall x_n$ may be written
as $\forall \bar{x}$.

The logic accords a special status to the various logical symbols that
are part of the language by including inference rules for interpreting
assumptions and for deriving formulas that contain them. The
interpretation of the logical symbols other than $=$ is similar
to the way we understand them in usual reasoning contexts. We will not
present these rules explicitly, but will use them in understandable
ways when we show derivations. The interpretation of equality is one
of the things that distinguishes Abella. The symbol $=$ is assumed to
have a fixed meaning in the logic: it is treated as $\beta
\eta$-convertibility. This interpretation does not seem remarkable
when it is applied to proving a formula with the equality symbol in
it; however, its unusual nature becomes clear when it is applied to an
equality assumption.  In this case, we would need to examine the
different ways in which the equality could hold and show that the
desired conclusion follows in all these cases. As a specific instance
of the use of this pattern of reasoning, assuming $a$ and $b$ to be
two distinct constants, the formula $a = b \supset \bot$ is
provable. This is because the two terms in the equality assumption are
not $\beta\eta$-convertible.

Another unusual aspect of the logic underlying Abella is that it
interprets atomic formulas using \emph{fixed-point definitions}.  Such
definitions are given by a collection of \emph{definitional clauses}
that have the form $\forall \bar{x}.(A \triangleq B)$, where $A$ is an
atomic formula with variables bound by $\bar{x}$ and $B$ is a formula.
The atomic formula $A$ is referred to as the head of the definition, and
$B$ is the body. The interpretation of a fixed-point definition is
that an atom $A$ holds if and only if $A$ matches with the head of an
instance of one of the clauses it contains and the body of the
corresponding clause holds. In writing clauses in Abella we typically
leave the universal quantifiers at the front implicit, showing the
variables they quantify by using symbols beginning with capital
letters.

Let us illustrate the ideas underlying the treatment of atomic
formulas in Abella by considering a definition of an ``append''
relation. To begin with, let $\mathsf{int}$ and $\mathsf{ilist}$ be atomic
types, and let $\mathsf{nil}: \mathsf{ilist}$ and
$\mathsf{cons}: \mathsf{int} \rightarrow \mathsf{ilist} \rightarrow \mathsf{ilist}$
be two constants that we use to construct
representations of lists of objects of type $\mathsf{int}$. Then we might
denote the append relation using the constant
\begin{center}
  $\mathsf{app}: \mathsf{ilist} \rightarrow \mathsf{ilist} \rightarrow \mathsf{ilist} \rightarrow \mathsf{prop}$
\end{center}
that is defined by the definitional clauses
\begin{center}
  $\appFour{\mathsf{app}}{\mathsf{nil}}{L}{L} \triangleq \top
  \qquad
  \appFour{\mathsf{app}}{(\appThree{\mathsf{cons}}{X}{L_1})}{L_2}{(\appThree{\mathsf{cons}}{X}{L_3})}
  \triangleq \appFour{\mathsf{app}}{L_1}{L_2}{L_3}$
\end{center}
One use for these clauses is to prove when the append relation
holds. As an example, consider the assertion
\[
\appFour{\mathsf{app}}
        {(\appThree{\mathsf{cons}}{1}{\mathsf{nil}})}
        {(\appThree{\mathsf{cons}}{2}{\mathsf{nil})}}
        {(\appThree{\mathsf{cons}}{1}{(\appThree{\mathsf{cons}}{2}{\mathsf{nil}})})}.
\]
This assertion is true if $\mathsf{app}$ is a predicate that is defined by
the clauses shown above. To actually construct a proof, we would match
the formula with the head of the second clause and ``unfold'' it into
the corresponding body that would then be proved by matching it with the
first clause. This process is similar in spirit to the one used to
construct proofs in the context of the HOHH logic that we saw in
Chapter~\ref{ch:hohh}. The difference between how clauses are
interpreted in Abella and in the HOHH logic shows up in the
case where we have the append relation appearing as an assumption in
proofs we want
to construct. As an example of this kind, consider the assertion
\[
\appFour{\mathsf{app}}{(\appThree{\mathsf{cons}}{1}{\mathsf{nil}})}{(\appThree{\mathsf{cons}}{2}{\mathsf{nil}})}{\mathsf{nil}} \supset \bot.
\]
In this case, we would want to show that if we assume
$\appFour{\mathsf{app}}{(\appThree{\mathsf{cons}}{1}{\mathsf{nil}})}{(\appThree{\mathsf{cons}}{2}{\mathsf{nil}})}{\mathsf{nil}}$
is true, then $\bot$ follows. Here
we make crucial use of the fact that an append
assumption can be true only because of one of the clauses defining
$\mathsf{app}$. This leads to a case analysis style of reasoning. We note
in this case that neither of the clauses for $\mathsf{app}$ could match
with the assumption we are claiming to be true and hence any
conclusion, including $\bot$, follows from it. Thus the assertion
under consideration has a proof in Abella.

Atomic formulas that are defined via fixed-point definitions can also
be reasoned about inductively in Abella. This style of reasoning
applies when we want to prove a formula of the form
\[\forall \bar{x}. F_1 \supset ... \supset A \supset ... \supset F_n \supset F_0,\]
where $A$ is defined by a fixed-point definition. Deciding that we
want to prove this formula by induction on $A$ gives us
the inductive hypothesis
\[\forall \bar{x}. F_1 \supset ... \supset A^* \supset ... \supset F_n \supset F_0\]
and it transforms the formula we want to prove into
\[\forall \bar{x}. F_1 \supset ... \supset A^{@} \supset ... \supset F_n \supset F_0.\]
The meaning of the ${@}$ and $*$ annotations is to be understood as
follows:  A formula with an ${@}$ annotation is considered ``larger
than'' a formula with a $*$ annotation but unfolding the former using
a definitional clause yields formulas with the $*$ annotation. Thus,
such a formula can match with the one in the induction hypothesis,
that is, this hypothesis can be used with the formula after it has been
unfolded.

The induction principle that we have described above is quite powerful
and can be used to prove a number of properties concerning predicates
described by fixed-point definitions. As an example, consider the
following formula that says that $\mathsf{app}$ is functional in its
behavior:
\[ \forall l_1 \forall l_2 \forall l_3 \forall l_4. \appFour{\mathsf{app}}{l_1}{l_2}{l_3} \supset
  \appFour{\mathsf{app}}{l_1}{l_2}{l_4} \supset l_3 = l_4.\]
If we try to prove this by only case analysis on the first or the
second assumption in this formula, we will get stuck in a cycle: the
second case in the definition of $\mathsf{app}$ will lead us back to trying
to prove a formula that has the same structure as the given one. This
situation reflects the fact that, in the general case, we do not know
the length of the list $l_1$ and hence are stuck with proving the same
formula, even if only for a shorter list. However, if we are able to
reason inductively on the definition of $\mathsf{app}$, we are able to
capture the effect of assuming that the formula we want to prove is
true when the list $l_1$ is of shorter length, and the proof then goes
through.

The treatment of atomic formulas has the consequence of giving
universal quantifiers an extensional interpretation. To see this,
suppose our definition is comprised of the following clauses
\[
   p\ a \triangleq \top \qquad q\ a \triangleq \top \qquad q\ b
   \triangleq \top
\]
and then consider the assertion $\forall x. (p\ x) \supset (q\ x)$.
This formula is provable, but the reason for this is that the only
thing of which $p$ is true, $a$, is such that $q$ is also true of
it. While this kind of quantification is often useful, sometimes we
also want to be able to show that a given formula has a \emph{generic}
proof, that is, the formula is true for the same reason for each
instance. To provide the ability to capture this notion, the logic
underlying Abella includes a new kind of quantifier, called a
\emph{nabla} quantifier. This quantifier is denoted by the symbol
$\nabla_{\tau} : (\tau \rightarrow \mathsf{prop}) \rightarrow
\mathsf{prop}$ that is pronounced ``nabla''; as with $\forall$ and
$\exists$, we drop types and also use a more suggestive ``quantifier''
notation when writing the $\nabla$ quantifier in formulas.  Now, to
prove a $\nabla$-quantified formula, we need to introduce a new
constant called a \emph{nominal constant} and then try to prove the
resulting instance. A key aspect about nominal constants is that their
structure is fixed; they cannot be further elaborated in the
course of constructing a proof. Thus, the proof we construct for
formulas involving the nabla quantifier has a generic structure of the
kind desired.

\section{Constructing Proofs}

Abella is used by issuing commands and using tactics.  To create a
fixed-point definition, the \verb|Define| command is used.  This takes
a name for the fixed-point definition, along with its type, and is
followed by the semicolon-separated definitional clauses.  If the body
of a definitional clause is simply $\top$, it may be omitted.  As an example,
the $\mathsf{app}$ predicate defined in the previous section would be encoded as

{\singlespacing
\begin{verbatim}
  Define app : ilist -> ilist -> ilist -> prop by
  app nil L L;
  app (cons X L1) L2 (cons X L3) := app L1 L2 L3.
\end{verbatim}
}


To start a proof, we declare a theorem using the \verb|Theorem| command,
which takes a name and the formula for the theorem.  For example,
suppose that we want to declare and prove the theorem about the
functional nature of append that we considered in the previous
section. We would get started on this by using the following
declaration in Abella:

{\singlespacing
\begin{verbatim}
   Theorem appFun : forall l1 l2 l3 l4,
        app l1 l2 l3 -> app l1 l2 l4 -> l3 = l4.
\end{verbatim}
}

\noindent Once we have declared a theorem, proving it becomes a
goal. Goals of this kind are presented as proof states by Abella. A
proof state consists of a collection of \emph{eigenvariables} that
represent universal quantifiers at the level of a proof, a set of
assumptions, and a formula that must be shown to be true in the context
of the assumptions. For example, after the theorem declaration above,
Abella will show us the following:

{\singlespacing
\begin{verbatim}

============================
 forall l1 l2 l3 l4, app l1 l2 l3 -> app l1 l2 l4 -> l3 = l4

appFun <
\end{verbatim}
}

\noindent Generally, the eigenvariables and assumptions are shown
above the line and the formula to be proven, the \emph{goal formula}
of the proof state, appears below. When we try to solve a particular
goal, this may spawn multiple subgoals, each of which will be
represented by a corresponding proof state. Abella will show us only the
first of these proof states in full detail; it will hold the others
for consideration after we have finished solving the subgoal that is
currently in focus.

To progress in the solution of a goal in this context, we use
tactics. One example of a tactic is that for using induction in the
form that we described it in the previous section. To construct a
proof by induction on the $i^{th}$ premise in an implicational
formula, we invoke this tactic through a command of the form
\verb|induction on i|. For example, in the proof state shown above, we
could invoke it as follows, leading to the new proof state that is
shown immediately after:

{\singlespacing
\begin{verbatim}
appFun < induction on 1.

IH : forall l1 l2 l3 l4, app l1 l2 l3 * -> app l1 l2 l4 -> l3 = l4
============================
 forall l1 l2 l3 l4, app l1 l2 l3 @ -> app l1 l2 l4 -> l3 = l4

appFun <
\end{verbatim}
}

\noindent Note that the hypotheses are identified by labels---the induction
hypothesis here has been given the label \verb+IH+. This is done so as
to enable us to name the particular hypotheses that we may want to use
in applying further tactics.

Continuing with our example, we might now want to simplify the goal
formula of the proof state using rules for introducing implications
and universal quantifiers. To do this we would invoke the
\verb+intros+ tactic, which leads to the following proof state:

{\singlespacing
\begin{verbatim}
Variables: l1 l2 l3 l4
IH : forall l1 l2 l3 l4, app l1 l2 l3 * -> app l1 l2 l4 -> l3 = l4
H1 : app l1 l2 l3 @
H2 : app l1 l2 l4
============================
 l3 = l4

appFun <
\end{verbatim}
}
\noindent All four universally-quantified variables were replaced by
eigenvariables, and these are shown at the top of the proof state.  We
also have the assumptions about the relations of appending lists $l_1$
and $l_2$.  Note that \verb|H1| has the ${@}$ annotation to mark it as
a larger version than is compatible with the inductive hypothesis.

The next step in completing the proof would be to do a case analysis
on one of the newly introduced goals. This is done using the
\verb+case+ tactic that takes as an argument an assumption formula,
indicated by its label:

{\singlespacing
\begin{verbatim}
appFun < case H1.
Subgoal 1:

Variables: l3 l4
IH : forall l1 l2 l3 l4, app l1 l2 l3 * -> app l1 l2 l4 -> l3 = l4
H2 : app nil l3 l4
============================
 l3 = l4

Subgoal 2 is:
 cons X L3 = l4

appFun <
\end{verbatim}
}

\noindent Observe that case analysis has resulted in two subgoals
here, only one of which is shown explicitly. This first subgoal can be
solved easily by using case analysis again on the assumption
\verb+H2+; this assumption can be true only because of the first
clause in the definition of \verb+app+, leading to the conclusion that
\verb+l3+ and \verb+l4+ must be equal in this case.






This leaves us with having to solve the second subgoal. Using case
analysis on the second \verb+app+ hypothesis leaves us in the
following state:

{\singlespacing
\begin{verbatim}
Subgoal 2:

Variables: l2 L3 X L1 L5
IH : forall l1 l2 l3 l4, app l1 l2 l3 * -> app l1 l2 l4 -> l3 = l4
H3 : app L1 l2 L3 *
H4 : app L1 l2 L5
============================
 cons X L3 = cons X L5

appFun <
\end{verbatim}
}

\noindent To complete this proof, we have to ``apply'' the induction
hypothesis to the other two assumptions. Abella provides an
\verb+apply+ tactic for this purpose. To use this tactic, we have to
identify a formula to be applied and the formula or formulas that it
should be applied to. These formulas can be hypotheses in the proof
state or previously proven theorems. In the present context, we can
invoke it as follows with the indicated result:

{\singlespacing
\begin{verbatim}
appFun < apply IH to H3 H4.
Subgoal 2:

Variables: L2 X L5 L9
IH : forall L1 L2 L3 L4, app L1 L2 L3 * -> app L1 L2 L4 -> L3 = L4
H3 : app L5 l2 L9 *
H4 : app L5 l2 L9
============================
 cons X L9 = cons X L9

appFun <
\end{verbatim}
}

\noindent The new proof state has a trivial proof since the goal
formula asserts equality between identical terms. Abella provides a
\verb+search+ tactic that can be invoked to try and complete proofs
that can be found with the application of a few simple steps. This
tactic can be used as the last step in this case.






Once all the subgoals have been proven, the proof is completed, and
the theorem can be used in future developments.

We have provided an exposure to some of the tactics available with
Abella through the example we have considered. These are not the only
tactics available, but they cover the ones we will use in the proofs we
consider in this thesis with one exception. Some of the theorems we
will want to prove will require the use of mutual induction. In such a
case, we will want to prove a conjunction of formulas. To do this, the
induction tactic allows us to identify a formula to do induction on in
each of the conjuncts. Subsequently, we may invoke the \verb|split|
tactic to break up the task into the subgoals of proving each of the
conjuncts separately.  Once a theorem that is a conjunction of formulas
has been proven this way, we can use the top-level command \verb|Split|
to separate the conjuncts into separate theorems that can be used
individually.

A more complete exposition of Abella may be found in \cite{abellaIntro}.

\section{The Encoding of HOHH in Abella}

Since the language of Abella is the same as that of the HOHH logic,
formulas of HOHH can be written more or less directly as Abella
formulas. We
can then encode the two kinds of sequents in HOHH using the predicates
$\mathsf{seq} : \mathsf{olist} \rightarrow \mathsf{o} \rightarrow
\mathsf{prop}$ and $\mathsf{foc} : \mathsf{olist} \rightarrow
\mathsf{o} \rightarrow \mathsf{o} \rightarrow \mathsf{prop}$; the
intention is that the first predicate should be defined so as to be
true exactly when it corresponds to a normal kind of sequent that is
derivable in the HOHH logic and the second should be defined so that
it is true when the corresponding focused sequent is derivable.
Note that the type $\mathsf{olist}$ corresponds to lists of formulas
in both cases, the last argument represents the formula on the
righthand side of a sequent and the middle argument in the case of
$\mathsf{foc}$ corresponds to the focus formula.

\begin{figure}
  \begin{center}
    \begin{tabular}{l}
      $\appThree{\mathsf{seq}}{L}{\top}                  \triangleq  \top$   \\
      $\appThree{\mathsf{seq}}{L}{(F \Rightarrow G)}     \triangleq  \appThree{\mathsf{seq}}{(F::L)}{G}$   \\
      $\appThree{\mathsf{seq}}{L}{(G_1\ \&\ G_2)}         \triangleq  \appThree{\mathsf{seq}}{L}{G_1} \wedge \appThree{\mathsf{seq}}{L}{G_2}$   \\
      $\appThree{\mathsf{seq}}{L}{(\Pi x:\tau. (F\ x))}  \triangleq  \nabla x:\tau. \appThree{\mathsf{seq}}{L}{(F\ x)}$   \\
      $\appThree{\mathsf{seq}}{L}{A}                     \triangleq  \appTwo{\mathsf{atom}}{A} \wedge \appThree{\mathsf{member}}{F}{L} \wedge \appFour{\mathsf{foc}}{L}{F}{A}$   \\
      \\
      $\appFour{\mathsf{foc}}{L}{(G \Rightarrow F)}{A}    \triangleq  \appThree{\mathsf{seq}}{L}{G} \wedge \appFour{\mathsf{foc}}{L}{F}{A}$ \\
      $\appFour{\mathsf{foc}}{L}{(\Pi x:\tau. (F\ x))}{A}  \triangleq  \exists t:\tau. \appFour{\mathsf{foc}}{L}{(F\ t)}{A}$ \\
      $\appFour{\mathsf{foc}}{L}{A}{A}                     \triangleq  \top$
    \end{tabular}
  \end{center}

  \caption{Encoding of inference rules of HOHH into the logic of Abella as fixed-point definitions.  In the clause for $\appFour{\mathsf{foc}}{L}{(F_1\ \&\ F_2)}{A}$, $i$ is either 1 or 2.}
  \label{fig:HOHHfixpoint}
\end{figure}

Figure \ref{fig:HOHHfixpoint} presents definitions in Abella for the
two predicates that implement their intended meanings. The clauses in
this figure are more or less transparent renditions of the inference
rules for HOHH. One thing to note is that $\nabla$ is used to encode
the $\Pi R$ rule.  This is because the interpretation of $\Pi$ in the
HOHH logic is that of a generic quantifier rather than that of the
universal quantifier in Abella. Perhaps the only things to be
explained are the predicates $\mathsf{atom}$ and $\mathsf{member}$ in
the last clause for $\mathsf{seq}$. The first predicate has the type
$\mathsf{o} \rightarrow \mathsf{prop}$ and is supposed to recognize
the encodings of atomic HOHH formulas. The second predicate checks for
the membership of a formula in a list of formulas. Both predicates can
be defined in Abella.



The encoding that we have described is, in fact, built into Abella to
give it the ability to reason about HOHH specifications. Abella also
has a special syntax for the encodings of the two forms of HOHH
sequents that we shall use.  A ``goal-reduction'' sequent of the form
$\appThree{\mathsf{seq}}{L}{G}$ is written as $\{L \vdash G\}$.
Backchaining sequents of the form $\appFour{\mathsf{foc}}{L}{F}{G}$
are written as $\{L, [F] \vdash G\}$.  When writing sequents using
this notation, if we wish to explicitly list several members of the
context, we shall include them in sequence, possibly after a schematic
variable denoting the rest of the list of assumptions as we see in the
following example: $\{L, p, q \vdash G\}$. In the case that the
assumption list is empty, the representation of goal-reduction
sequents is simplified to $\{ G\}$.

\section{Reasoning About HOHH Specifications} \label{sec:reasoningHOHH}

Once we have specified a rule-based system in HOHH, we can reason
about that specification in Abella.  This is done in the same way
as reasoning about definitions created directly in Abella, since
the rules of HOHH are encoded as a fixed-point definition.  Then
we are able to work with them just as with any other definition.

As an example, we show that appending two lists is deterministic
using the rules for appending lists written in HOHH in Section
\ref{sec:encodingRules},
\[\appFour{\mathsf{append}} {\mathsf{nil}} {L} {L}
        \qquad
  \appFour{\mathsf{append}} {L_1} {L_2} {L_3} \Rightarrow
        \appFour{\mathsf{append}} {(\appThree{\mathsf{cons}}{X}{L_1})} {L_2}
             {(\appThree{\mathsf{cons}} {X} {L_3})}\]
We can write a theorem $\mathsf{app\_determ}$
to show that $\mathsf{append}$ is deterministic:
\[\forall L_1 \forall L_2 \forall L_3 \forall L_4.
    \{\appFour{\mathsf{append}}{L_1}{L_2}{L_3}\} \supset
    \{\appFour{\mathsf{append}}{L_1}{L_2}{L_4}\} \supset L_3 = L_4\]
The meaning of this theorem is that if we have a derivation of
$\appFour{\mathsf{append}}{L_1}{L_2}{L_3}$ in HOHH in the context
of the given rules for $\mathsf{append}$ and a derivation of
$\appFour{\mathsf{append}}{L_1}{L_2}{L_4}$ in the same context, then
the two lists $L_3$ and $L_4$ must, in fact, be the same list.
We prove this by induction on the first sequent.
The induction tactic creates an inductive hypothesis
\[\forall L_1 \forall L_2 \forall L_3 \forall L_4.
    \{\appFour{\mathsf{append}}{L_1}{L_2}{L_3}\}^* \supset
    \{\appFour{\mathsf{append}}{L_1}{L_2}{L_4}\} \supset L_3 = L_4\]
Note the annotation on the first $\mathsf{append}$ sequent, marking
it to only be used by a strictly smaller instance of a hypothesis
matching it.  A strictly smaller instance of a sequent is actually a
shorter derivation of a similar goal formula, rather than directly
referring to using $\mathsf{append}$ on a shorter list, since we
are reasoning about the HOHH derivation rather than the
$\mathsf{append}$ relation itself.

By application of the \verb|intros| tactic we get
the hypotheses $\mathsf{H1}: \{\mathsf{append}\ L_1\ L_2\ L_3\}^@$ and
$\mathsf{H2}: \{\mathsf{append}\ L_1\ L_2\ L_4\}$, as well as adding the
eigenvariables $L_1, L_2, L_3,$ and $L_4$ into the context, and the
conclusion we wish to reach is now reduced to $L_3 = L_4$.  We analyze
the possible cases for $\mathsf{H1}$ using the $\mathsf{case}$ tactic.
The cases of this are the inference rules that could have been used to
prove it.  Since the goal is atomic, the cases are the formulas that
may be focused on and used, which are the two formulas for
$\mathsf{append}$, app-nil and app-cons.

If $\mathsf{H1}$ holds by focusing on the app-nil rule, $L_1$
must be $\mathsf{nil}$ and $L_2 = L_3$.  There is no shorter
derivation involved with this, since it is solved by \init,
so we get no new hypotheses.  Then, by using the equality of
$L_2$ and $L_3$, $\mathsf{H2}$ is automatically transformed to
$\{\appFour{\mathsf{append}}{\mathsf{nil}}{L_3}{L_4}\}$.
Doing case analysis on this hypothesis, since only the app-nil
rule can be focused on, shows that $L_3 = L_4$, finishing the subgoal.

Alternatively, if $\mathsf{H1}$ holds by focusing on and using
the app-cons rule for backchaining, the list $L_1$
must be the result of constructing a list from a list element $X$
and a shorter list $L_1'$.  Then $L_3 = X::L_3'$, for some $L_3'$,
to match the app-cons rule.  $\mathsf{H1}$ is replaced by
$\mathsf{H3}: \{\appFour{\mathsf{append}}{L_1'}{L_2}{L_3'}\}^*$, since
focusing on app-cons in a derivation leads to having to show that the
antecedent of the clause holds as well, and then we may assume that
this antecedent holds when the formula has been successfully used in
a derivation.  Note that
the annotation on $\mathsf{H3}$ marks it as a shorter derivation
which can be used with the inductive hypothesis.  The hypothesis $\mathsf{H2}$
is transformed to $\{\appFour{\mathsf{append}}{(X::L_1')}{L_2}{L_4}\}$.
Carrying out case analysis on this, we can only backchain on app-cons
since the first list
is non-empty, and this tells us that $L_4 = X::L_4'$, for some list $L_4'$,
creating the new hypothesis
$\mathsf{H4}: \{\appFour{\mathsf{append}}{L_1'}{L_2}{L_4'}\}$.  The inductive
hypothesis can be applied to $\mathsf{H4}$ and $\mathsf{H5}$, which shows that
$L_3' = L_4'$.  Since our goal is $X::L_3' = X::L_4'$, this result leads to our
goal being rewritten as $X::L_4' = X::L_4'$, and it can be solved with the
\verb|search| tactic.  Since this was the last subgoal, the proof is completed
and $\mathsf{app\_determ}$ is added as a theorem that can be applied in future
proofs.

%% file: strengthening.tex
\chapter{Proving Strengthening Lemmas}\label{ch:strengthening-proof}

This chapter describes how strengthening lemmas are proven in
Abella. As we have seen in Chapter \ref{ch:hohh}, the dynamic context
can grow when we are trying to construct a proof for a goal formula in
the HOHH logic. For this reason, we have to first generalize the
strengthening lemma to take into account the different forms the
dynamic context can have. A second aspect to pay attention to is that
the proof of a particular goal formula may depend on the provability
of additional goal formulas---these are the ``subgoals'' that arise in
constructing a proof in the HOHH logic. Thus, we may have to prove
additional strengthening lemmas for these other goal formulas and all
these proofs will have to be constructed simultaneously using mutual
induction. 

In Chapter~\ref{ch:dependencies} we have examined how we can determine
if a strengthening lemma actually holds. The information we calculated
there provides us a means for determining the structure of the dynamic
contexts and the goals that arise in the course of a proof. In this
chapter we describe how this information can be converted into a form
from which an explicit proof can be generated for a given strengthening
lemma in Abella. In the first section below, we consider the construction of
definitions for the dynamic contexts. We then describe how to generate
a strengthened form of the strengthening lemma. The last two sections
consider in turn the automatic generation of a proof in Abella for the
stronger theorem and the incorporation of these ideas in the form of a
new tactic in Abella for automating the proof of strengthening lemmas.

\section{Formalizing Dynamic Contexts} \label{sec:dynCtxDefinitions}

To start, the formulas that may dynamically appear in the contexts of
each of the predicates are calculated, as discussed in Section
\ref{sec:dynCtx}.  The dynamic context of a predicate is then defined
by a fixed-point definition in the logic of Abella defining a list that either
contains nothing, or contains only the formulas from the dynamic
context of the predicate.  Then it has definitional clauses for
$\mathsf{nil}$ and for each formula of the dynamic context.  The rest
of this work will assume that the dynamic context predicate for a
predicate $p$ is named $\ctx{p}$.  If the dynamic context of a
predicate $p$ may contain formulas $F_1, ..., F_n$, then we have the
definitional clauses 
\[\appTwo{\ctx{p}}{\mathsf{nil}} \triangleq \top \qquad
  \appTwo{\ctx{p}}{(F_1::L)} \triangleq \appTwo{\ctx{p}}{L} \qquad ... \qquad
  \appTwo{\ctx{p}}{(F_n::L)} \triangleq \appTwo{\ctx{p}}{L}\]
Such a definition is created for the dynamic context of each predicate
the goal that is being strengthened depends on, as calculated by the
algorithm discussed in Section \ref{sec:dependencies}. 

To assist in the proof of the strengthening lemma, a lemma is created
and proven that shows if a list is of the form of the dynamic context
for some predicate, then a member of it must be one of a limited
number of forms, specifically the formulas that might occur in the
dynamic context of that predicate.  We refer to this as a context
membership lemma, and assume the name of the context membership lemma
for a predicate $p$ is $\ctxmem{p}$.  If there are no formulas that
may appear in the dynamic context, it is asserted that having a member
of such a list leads to a contradiction.  Then the theorem for the 
predicate $p$ is of the form 
\[\forall E \forall L. \appTwo{\ctx{p}}{L} \supset 
   \appThree{\mathsf{member}}{E}{L} \supset 
    \exists \bar{x}_1.E = F_1 \vee ... \vee \exists \bar{x}_n.E = F_n\]
where the dynamic context of $p$ may contain the formulas
$F_1,...,F_n$ and $\bar{x}_i$ contains the variables in $F_i$.

\begin{figure}
  \begin{center}
    \begin{minipage}{2.75in}
      \begin{tabbing}
        \qquad \= \kill
          \verb+induction on 1. intros. case H1.+\\
          \verb+case H2.+\\
          \textbf{for} $i = 1..n$\\
        \>   \verb|case H2. search.|\\
        \>   \verb|apply IH to H3 H4. search.|\\
          \textbf{end for}
      \end{tabbing}
    \end{minipage}
  \end{center}
  \caption{A structure for the proof of the context membership lemma for
    a predicate $p$.  The assumption is that the dynamic context of
    $p$ contains $n$ formulas where $n \geq 0$.} 
  \label{fig:ctxMemPrf}
\end{figure}

Figure \ref{fig:ctxMemPrf} shows the form for the proof of the
context membership lemmas that are generated; we mean the for loop
here to be read as a means for showing the (static) repetition of the
tactic invocations in the body of the loop and not as a higher-order
form of tactic that can be invoked dynamically. The structure of the
proof that is constructed can be understood as follows:  To prove this
formula, we carry out induction on the fixed-point definition of the
context.  After declaring our induction and introducing the
hypotheses, we have eigenvariables $E$ and $L$ and hypotheses
$\mathsf{H1}: \appTwo{\ctx{p}}{L}$ and
$\mathsf{H2}: \appThree{\mathsf{member}}{E}{\mathsf{nil}}$, and our
goal is
$\exists \bar{x}_1.E = F_1 \vee ... \vee \exists \bar{x}_n.E = F_n$.
We carry out case analysis on the hypothesis corresponding to our
induction (\verb|case H1|).  After this, we have the single hypothesis
$\mathsf{H2}$ remaining.  Since it is not possible to have a member of
an empty list, case analysis on this solves the current subgoal.
Then, for each formula $F_i$ that can appear in the dynamic context,
we have two hypotheses,
\begin{center}
  \renewcommand{\arraystretch}{.7}
  \begin{tabular}{ll}
    $\mathsf{H2}: \appThree{\mathsf{member}}{E}{(F_i::L)}$\\
    $\mathsf{H3}: \appTwo{\ctx{p}}{L}$
  \end{tabular}
\end{center}
and the same goal as before.  Case analysis on $\mathsf{H2}$ gives us
that $E = F_i$, and so \verb|search| will solve this subgoal. The next
subgoal is to show that, if $E$ is in the rest of the list rather than
being the first element, it is also one of the formulas of the dynamic
context.  This is done by application of the inductive hypothesis.
Carrying this out for all formulas that may be in the dynamic context
solves all subgoals, completing the proof.

For each pair of predicates $(a,b)$, where $a$ depends on $b$, a
subcontext lemma is generated.  This will simplify the proof of the
strengthening lemma.  Since all the formulas that may occur in $a$'s
dynamic context also occur in $b$'s dynamic context, it is the case
that an instance of $a$'s dynamic context is also an instance of $b$'s
dynamic context.  Then the subcontext theorem is 
\[\forall L. \appTwo{\ctx{a}}{L} \supset \appTwo{\ctx{b}}{L}\]
Hereafter we shall assume the subcontext theorem as stated above is
named $\subctx{a}{b}$. 

\begin{figure}
  \begin{center}
    \begin{minipage}{2.75in}
      \begin{tabbing}
        \qquad \= \kill
          \verb|induction on 1. intros. case H1.|\\
          \verb|search.|\\
          \textbf{for} $i = 1..n$\\
        \>   \verb|apply IH to H2. search.|\\
          \textbf{end for}
      \end{tabbing}
    \end{minipage}
  \end{center}

  \caption{Structure of the proof for proving the subcontext
    relationship where the dynamic context of $a$ is a subset of the
    dynamic context of $b$, where $a$'s dynamic context contains $n$
    formulas with $n \geq 0$.}
  \label{fig:subctxPrf}
\end{figure}

The proof structure for subcontext lemmas is found in Figure
\ref{fig:subctxPrf}.  The proof is done by induction on the assumption
that the list is of the form of an instance of $a$'s dynamic context.
The \verb|intros| tactic introduces this as an assumption
$\mathsf{H1}$, leaving the goal $\appTwo{\ctx{b}}{L}$, and case
analysis on $\mathsf{H1}$ allows us to go through each possible form
of the list.  The first possibility is the empty list, which is solved
simply by using the \verb|search| tactic, since the goal becomes
$\appTwo{\ctx{b}}{\mathsf{nil}}$, and $\ctx{b}$ is also defined to fit
the empty list.  As before, the for loop represents repetition of the
body rather than a tactic.  For each subsequent formula $F_i$, the goal
formula is $\appTwo{\ctx{b}}{(F_i::L)}$, with the single hypothesis
$\mathsf{H2}: \appTwo{\ctx{a}}{L}^*$.  Applying the inductive hypothesis
shows that it is also the case that $L$ is an instance of $\ctx{b}$.
With this information, \verb|search| is able to find the definitional
clause for such a list for $\ctx{b}$ and finish the proof of the
subgoal.

\section{Generation of Strengthening Lemmas}

To strengthen a formula $G$ from its dependence on a formula $F$, it
must be shown that, for every situation that may arise in proving $G$,
$F$ cannot be used.  It must be further shown for each goal that may
arise, no matter which formulas are in its dynamic context, $F$ cannot
be used; that is, to strengthen $G$, we must strengthen not only $G$
but also all the predicates its head predicate depends on from $F$.
These cannot be carried out entirely separately in all cases, however;
in some cases they must be carried out through mutual induction for
all the predicates $G$ depends on.  This allows an inductive
hypothesis to be used to show, when a goal $P$, where $\headPred{P} =
p$, backchains to create a goal $Q$, where $\headPred{Q} = q$, that
$F$ will still not be used in the proof of $Q$, and so will not be
used in the proof of $P$.  If $q$ does not depend on $p$, then proving
the strengthening lemma for $q$ first and simply using it in the proof
of the strengthening lemma for $p$ will work; however, it is possible
for them to be mutually dependent.  Then neither one can be proven
without appealing to the strengthening of the other one.  This is the
kind of situation which requires that the proof be made with mutual
induction.  It is also possible to have larger loops of dependence,
such as having $p$ depend on $q$, which depends on $r$, and $r$
depends on $p$ again.  In this case as well, none of the predicates'
strengthening lemmas can be proven without appealing to another's.

Each predicate may have a different set of formulas that can appear in
its dynamic context.  Then for each predicate we show its
strengthening from its own dynamic context.  The possibility must be
considered that a formula that appears in the dynamic context might be
used in the proof, and that this might lead to using the formula that
is being strengthened from.  We cannot use one context that contains
all formulas that may arise in the dynamic contexts of all predicates
being strengthened either, without the possibility of severely
limiting the strengthening lemmas that may be proven.  Consider the
case where we have two predicates $p$ and $q$, where $q$'s dynamic
context contains $F \Rightarrow p$, but this formula does not appear
in $p$'s dynamic context.  Then using one overarching context
definition would lead to this being available when proving $p$, and so
it would appear $F$ could be used in the proof of $p$ when it actually
could not.  If there are no formulas like this that would interfere
with the ability to prove strengthening, then having a single context
would work; however, this is not the general case, and so we use
separate dynamic context definitions for each predicate.

The form of a strengthening lemma for a predicate $p$  where we are
strengthening from a formula $F$ is
\[\forall L \forall \bar{x}. \appTwo{\ctx{p}}{L} \supset
         \{L, F \vdash G\} \supset \{L \vdash G\}\]
where $\headPred{G} = p$ and $\bar{x}$ contains all
universally-quantified variables that appear in $F$ and $G$.  Proving
this shows that having a proof of $G$ from a list $L$ representing an
instance of the dynamic context of $p$ and a formula $F$ means that a
proof of $G$ can also be derived from $L$ alone.  If we have a set of
predicates $p_1, ..., p_n$ to strengthen from $F$, we write the
mutually-inductive strengthening lemma as the conjunction of the
separate strengthening lemmas for all the predicates:
\begin{gather*}
  (\forall L \forall \bar{x}_1. \appTwo{\ctx{p_1}}{L} \supset
           \{L, F \vdash G_1\} \supset \{L \vdash G_1\}) \wedge
  ... \\ \qquad \wedge
  (\forall L \forall \bar{x}_n. \appTwo{\ctx{p_n}}{L} \supset
           \{L, F \vdash G_n\} \supset \{L \vdash G_n\})
\end{gather*}
where $\headPred{G_i} = p_i$.

\section{Generating Proofs for the Strengthening Lemmas}
           \label{sec:strenPrf}

\begin{figure}
  \begin{center}
    \begin{minipage}{5in}
      \begin{tabbing}
        \qquad \= \qquad \= \qquad \= \qquad \= \kill
          \verb|induction on 2|$*n$.\\ 
          \textbf{if} $n \geq 2$ \textbf{then}\\
        \>   \verb|split.|\\
          \textbf{end if}\\
          \textbf{for} $i = 1..n$\\
        \>   \verb|intros. case H2.|\\
        \>   \textbf{for} $D = (\Pi \bar{x}.(G_1\ \&\ ...\ \&\ G_m)
                   \Rightarrow A)$ where $\headPred{A} = a_i$ and $D
                   \in \Gamma$\\ 
        \> \>   \textbf{for} $j = 1..m$\\
        \> \> \>     \verb|apply |$\subctx{a_i}{\headPred{G_j}}$
                          \verb|to H1|.\\ 
        \> \> \>     \verb|apply IH|$_{\headPred{G_J}}$
                          \verb|to H|$(5+j+m)$
                          \verb|H|$(5+j)$.\\ 
        \> \>   \textbf{end for}\\
        \> \>   \verb|search.|\\
        \>   \textbf{end for}\\
        \>   \verb|case H4. case H3.|\\
        \>   \verb|apply |$\ctxmem{a_i}$\verb| to H1 H5.|\\
        \>   // dynamic context of $a_i = \{D_1,...,D_p\}$\\
        \>   \textbf{if} $p > 1$ \textbf{then}\\
        \> \>   \verb|case H6.|\\ 
        \>   \textbf{end if}\\
        \>   \textbf{for} $j = 1..p$\\
        \> \>   // $D_j = (\Pi \bar{x}.(G_1\ \&\ ...\ \&\ G_m)
                     \Rightarrow A)$ \\ 
        \> \>   \verb|case H3.|\\
        \> \>   \textbf{if} $\headPred{A} = a_i$ \textbf{then}\\
        \> \> \>     \textbf{for} $k = 1..m$\\
        \> \> \> \>       \verb|apply |$\subctx{a_i}{\headPred{G_k}}$
                               \verb|to H1|.\\
        \> \> \> \>       \verb|apply IH|$_{\headPred{G_k}}$
                               \verb|to H|$(5+k+m)$
                               \verb|H|$(5 + k)$.\\ 
        \> \> \>     \textbf{end for}\\
        \> \> \>     \verb|search.|\\
        \> \>   \textbf{end if}\\
        \>   \textbf{end for}\\
          \textbf{end for}
      \end{tabbing}
    \end{minipage}
  \end{center}

  \caption{Structure of the proof for the mutually-inductive
    strengthening lemma for a set of predicate dependencies
    $\{a_1,...,a_n\}$.  In the first line \texttt{2}$*n$ means $n$
    digit 2's.} 
  \label{fig:strenPrf}
\end{figure}

Once the mutually-inductive strengthening lemma has been generated, it
can be automatically proven.  We assume that the formula we wish to
strengthen from is $F$, and that the goal formula we ultimately wish
to strengthen depends on predicates $a_1, ..., a_n$.  For each
predicate $a_i$ we create a goal formula for the predicate by creating
universally-quantified variables for each of its arguments.  We call
the goal formula created in this way $A_i$.  The structure of the
proof for the mutually-inductive strengthening lemma for these
predicates is found in Figure \ref{fig:strenPrf}.  This proof is done
by induction on the unstrengthened derivations for each predicate.  We
refer to the inductive hypothesis for the predicate $a_i$ as
\verb|IH|$_{a_i}$.  Each predicate $a_i$'s strengthening is proven
separately, with the separation done by using the \verb|split| tactic
if we have more than one predicate.  For each predicate, we use the
\verb|intros| tactic to introduce eigenvariables and create hypotheses 
\begin{center}
  \renewcommand{\arraystretch}{.7}
  \begin{tabular}{ll}
    $\mathsf{H1}:$ & $\appTwo{\ctx{a_i}}{L}$ \\
    $\mathsf{H2}:$ & $\{L, F \vdash A_i\}$
  \end{tabular}
\end{center}
which leaves us with the goal $\{L \vdash A_i\}$.  We carry out case
analysis on $\mathsf{H2}$, which considers the cases for how
$\mathsf{H2}$ holds, whether by backchaining on a formula from the
static context or the dynamic context.

We start by iterating over the static context to find program clauses
that might be used as the last step in the derivation of $A_i$.  Any
program clause $D$ where $\headPred{D} \neq a_i$ is automatically
skipped as it is not possible for it to be used as the last step of
the derivation of $A_i$ in the inference rules of HOHH.  For any
program clause with the head predicate $a_i$, we backchain, which
creates assumptions for the derivations of all the antecedents, each
of which has the form $\{L, F \vdash G_j\}^*$.  For each of these
antecedents, the appropriate subcontext lemma is applied to show that
the current dynamic context is also an instance of the dynamic context
for $\headPred{G_j}$, and then the inductive hypothesis for
$\headPred{G_j}$ can be used to show that all derivations do not use
the formula being strengthened from.  This gives us a hypothesis $\{L
\vdash G_j\}$.  Once this is done for all the antecedents, the
\verb|search| tactic will finish proving that, when backchaining on
the current program clause, the derivation does not use the formula
being strengthened from.

Once this has been done for every program clause, the proof moves on
to attempting to use the dynamic context to backchain on, which
includes both the formula being strengthened from and the defined
dynamic context. Then we have the following hypotheses:
\begin{center}
  \renewcommand{\arraystretch}{.7}
  \begin{tabular}{ll}
    $\mathsf{H1}:$ & $\appTwo{\ctx{a_i}}{L}$ \\
    $\mathsf{H3}:$ & $\{L, F, [E] \vdash A_i\}^*$ \\
    $\mathsf{H4}:$ & $\appThree{\mathsf{member}}{E}{(F::L)}$
  \end{tabular}
\end{center}
We do case analysis on $\mathsf{H4}$, which gives us subgoals for
showing $\{L \vdash A_i\}$ in the cases where $E = F$ and where $E$ is
a member of the rest of $L$.  The case where $E = F$ is solved by case
analysis on $\mathsf{H3}$, since $F$ cannot be instantiated to match
$A_i$.  The context membership lemma for $a_i$ is then used to get the
cases for membership in the rest of the list, and the application of
this lemma creates a new hypothesis.  If there are multiple formulas
that may appear in the dynamic context, case analysis is done on this
hypothesis, and a subgoal is generated for each formula that may be a
member of the dynamic context.

Iterating through these formulas is very similar to iterating through
the static context formulas.  If a formula cannot be used to directly
solve the current goal, doing case analysis on $\mathsf{H3}$, the
unstrengthened derivation hypothesis, will immediately solve the goal.
If the current formula can be used to solve the current goal, it is
backchained on and we get hypotheses for each of the antecedents.  As
before, we use the appropriate subcontext lemma and inductive
hypothesis for each, then the \verb|search| tactic at the end to
finish the proof for the current dynamic context formula.

Once all the subgoals for the dynamic context formulas are finished,
we move on to the strengthening lemma for the next predicate and
repeat the process.  After the portion of the proof for the dynamic
context of the last predicate is finished, the whole
mutually-inductive strengthening lemma has been proven and can be
split and used in further developments.

\section{A Tactic for Proving Strengthening Lemmas}

Using the automatic generation of proofs of strengthening lemmas
discussed in the previous section, a tactic to automatically prove
strengthening has been implemented in Abella.  To use this tactic, a
user creates a fixed-point definition for a predicate defining a
context containing formulas $F_1,...,F_n$, where $n \geq 0$.  The user
then declares a theorem in the form of a strengthening lemma using
this context, which has the form 
\[\forall L \forall \bar{x}. \appTwo{ctx}{L} \supset
         \{L, F \vdash G\} \supset \{L \vdash G\}\]
where $F$ is the formula to be strengthened from, $\bar{x}$ contains
any quantified variables found in $F$ and $G$, and the name of the
defined context is $ctx$.  After this, he invokes the
\verb|strengthen| tactic. 

The \verb|strengthen| tactic adds the formulas $F_1,...,F_n$ to the
static context for calculating the dynamic contexts as discussed in
Section \ref{sec:dynCtx}, but also adds $F_1,...,F_n$ to each
predicate's dynamic context.  It does the same with any antecedents of
$G$.  These need to be part of the dynamic context for $\headPred{G}$,
since they are available for use in the derivation of $G$.  Then,
since they can be in the dynamic context of $\headPred{G}$, they must
also be part of the dynamic context for each predicate $\headPred{G}$
depends on.  After the dynamic contexts are calculated, the
dependencies are calculated as well, as discussed in Section
\ref{sec:dependencies}.  If $\headPred{F} \in S(\headPred{G})$, then
an error is thrown, since there may be a dependency between $G$ and
$F$, and the automated proof of the strengthening lemma cannot
succeed.

Once the dependencies are known, the dynamic context definition and
associated lemmas discussed in Section \ref{sec:dynCtxDefinitions} can
be defined and proven.  Using these, a mutually-inductive
strengthening lemma is created and proven, using the algorithm
discussed in Section \ref{sec:strenPrf}.  Once this proof is finished,
a subcontext lemma is proven to show that the user-defined context can
only contain a subset of the formulas that may occur in the calculated
dynamic context of $\headPred{G}$.  It can be seen that this is true,
since all the formulas of the user-defined context are included
automatically in the contexts that are automatically defined to create
the mutually-inductive strengthening lemma.  After being proven, the
automatically-generated mutually-inductive strengthening lemma is
split into its separate components using the \verb|Split| command.
Then the original theorem entered by the user is proven by using the
subcontext lemma for the user-created context definition and applying
the split strengthening lemma, with this proof shown in Figure
\ref{fig:strenThmPrf}.

\begin{figure}
  \begin{center}
    \begin{minipage}{3in}
      \verb|intros.|\\
      \verb|apply| $ctx\mathsf{\_subctx\_}\ctx{\headPred{G}}$.\\
      \verb|apply| \textit{split strengthening lemma} \verb|to H3 H2|.\\
      \verb|search.|
    \end{minipage}
  \end{center}

  \caption{Proof of the user-entered strengthening theorem, where
    $ctx$ is the name of the user-defined context, $G$ is the goal to
    be strengthened, and \textit{split strengthening lemma} refers to
    the portion of the mutually-inductive strengthening lemma for the
    predicate $\headPred{G}$.}
  \label{fig:strenThmPrf}
\end{figure}

It is necessary to run these proofs rather than just assume that
strengthening holds when $\headPred{G}$ does not depend on
$\headPred{F}$ in order to reduce the trusted code base of the proof
assistant.  If we trusted that the dependency calculations are
correct, then an error in them could invalidate any development using
the strengthen tactic.  By running the proofs explicitly, only the
other, lower-level tactics are trusted code, as is the case in the
proof assistant in general.  By keeping the trusted code base as small
as possible, the proof assistant is more trustworthy, as the small
trusted code base can be more easily verified than a larger code base
could be.

%% file: conclusion.tex
\chapter{Conclusion}

This thesis has shown how strengthening lemmas can be automatically generated
and proven in the Abella Proof Assistant.  This is done by carrying out a
reachability analysis to determine which types of formulas can arise in the
derivation of a given goal.  The analysis can then be used to generate
strengthening lemmas and explicit proofs of them.  It further describes how
automatically generating and proving strengthening lemmas can be used to implement
a \verb|strengthen| tactic that allows a user to create a simple strengthening
lemma and have the more complex background work and proof done for him.

The work in this thesis can be extended in several ways. We describe
two particular directions that look especially promising and that we
intend to explore in the future. In the first direction, we would like
to consider applications for the strengthening lemmas whose proofs we
have provided a means for automating. The immediate motivation for
considering these lemmas is that they enable the discovery that terms
of a particular type could not contain terms of another type, leading
thereby to the pruning of some branches in a case analysis over
equality assumptions. Now that we have a means for proving these
lemmas automatically, we would like to see how the process of using
them in the manner we have described can also be automated. The second
avenue for future work concerns the development of an algorithm that
provides a more careful analysis of dependencies and thereby enables
the validation of more strengthening lemmas. Specifically, we have
described in Section~\ref{sec:compConserv} how the current algorithm
misses some cases and we have also explained how it might be modified
to do better in these cases. We intend to both articulate an improved
analysis based on these ideas and also to evaluate whether the
additional strengthening lemmas it allows us to prove are an adequate
compensation for the more complex form to the generated proofs.

%% file: root.bbl
\begin{thebibliography}{1}

\bibitem{abellaIntro}
David Baelde, Kaustuv Chaudhuri, Andrew Gacek, Dale Miller, Gopalan Nadathur,
  Alwen Tiu, and Yuting Wang.
\newblock Abella: A system for reasoning about relational specifications.
\newblock {\em Journal of Formalized Reasoning}, 7, 2014.

\bibitem{stlcChurch}
Alonzo Church.
\newblock A formulation of the simple theory of types.
\newblock {\em Journal of Symbolic Logic}, 5:56 -- 68, 1940.

\bibitem{miller12proghol}
Dale Miller and Gopalan Nadathur.
\newblock {\em Programming with Higher-Order Logic}.
\newblock Cambridge University Press, June 2012.

\bibitem{uniformProofs}
Dale Miller, Gopalan Nadathur, Frank Pfenning, and Andre Scedrov.
\newblock Uniform proofs as a foundation for logic programming.
\newblock {\em Annals of Pure and Applied Logic}, 51:125--157, 1991.

\bibitem{teyjus.website}
Xiaochu Qi, Andrew Gacek, Steven Holte, Gopalan Nadathur, and Zach Snow.
\newblock The {T}eyjus system -- version 2, 2009.
\newblock http://teyjus.cs.umn.edu/.

\bibitem{yutingPaper}
Yuting Wang and Kaustuv Chaudhuri.
\newblock A proof-theoretic characterization of independence in type theory.
\newblock In Thorsten Altenkirch, editor, {\em Proceedings of the 13th
  International Conference on Typed Lambda Calculi and Applications (TLCA)},
  pages 332--346, Warsaw, Poland, July 2015.

\end{thebibliography}
